\documentclass[aps, pre, reprint,superscriptaddress]{revtex4-1}

\usepackage{graphicx}
\usepackage{dcolumn}
\usepackage{bm}

\PassOptionsToPackage{hyphens}{url}
\usepackage[
unicode=true,
 bookmarks=true,
 bookmarksnumbered=true,
  bookmarksopen=true,
 bookmarksopenlevel=2,
 bookmarksdepth=paragraph,
 breaklinks=false,
 pdfborder={0 0 0},
 backref=false, 
 colorlinks=false,
 allcolors=blue
 ]
 {hyperref}

\usepackage{booktabs}
\newcommand{\ra}[1]{\renewcommand{\arraystretch}{#1}}
\usepackage{setspace}
\usepackage{amsmath}

\usepackage{mathtools}
\usepackage{amssymb}
\usepackage{chngcntr}

\usepackage{todonotes}

\begin{document}

\title{Microbial range expansions on liquid substrates}

\author{Severine Atis}
\email[Severine Atis and Bryan T.\@ Weinstein contributed equally to this work:
]{severine.atis@gmail.com}
\affiliation{Department of Physics, Harvard University}

\author{Bryan T. Weinstein}
\email[Severine Atis and Bryan T.\@ Weinstein contributed equally to this work:
]{severine.atis@gmail.com}
\affiliation{School of Engineering and Applied Sciences, Harvard University}

\author{Andrew W. Murray}
\affiliation{FAS Center for Systems Biology, Harvard University}
\affiliation{Department of Molecular and Cellular Biology, Harvard University}

\author{David R. Nelson}
\affiliation{Department of Physics, Harvard University}
\affiliation{School of Engineering and Applied Sciences, Harvard University}
\affiliation{FAS Center for Systems Biology, Harvard University}
\affiliation{Department of Molecular and Cellular Biology, Harvard University}

\date{\today}

\begin{abstract}
Despite the importance that fluid flow plays in transporting and organizing populations, few laboratory systems exist to systematically investigate the impact of advection on their spatial evolutionary dynamics. To address this problem, we study the  morphology and genetic spatial structure of microbial colonies growing on the surface of a nutrient-laden fluid $10^4$ to $10^5$ times more viscous than water in Petri dishes; the extreme but finite viscosity inhibits undesired thermal convection and allows populations to effectively live at the air-liquid interface due to capillary forces. We discover that  \textit{S. cerevisiae} (baker's yeast) growing on a viscous liquid behave like ``active matter": they metabolically generate fluid flows many times larger than their unperturbed colony expansion speed, and that flow, in return, can dramatically impact their colony morphology and spatial population genetics. We show that yeast cells generate fluid flows by consuming surrounding nutrients and decreasing the local substrate density, leading to misaligned fluid pressure and density contours, which ultimately generates vorticity via a thresholdless baroclinic instability. Numerical simulations with experimentally measured parameters demonstrate that an intense vortex ring is produced below the colony's edge and quantitatively predict the observed flow. As the viscosity of the substrate is lowered and the self-induced flow intensifies, we observe three distinct morphologies: at the highest viscosity, cell proliferation and movement produces compact circular colonies similar to those grown on hard agar plates except with a stretched regime of exponential expansion, intermediate viscosities give rise to  compact colonies with ``fingers'' that are usually monoclonal and are ripped away to break into smaller cell clusters, and at the lowest viscosity, the expanding colony breaks up into many genetically-diverse, mutually repelling, island-like fragments of yeast colonies that can colonize an entire 94 mm-diameter Petri dish within 36 hours. We propose a simple phenomenological model in the spirit of the lubrication approximation that predicts the early colony dynamics. Our results provide rich opportunities for future investigations and suggest that microbial range expansions on viscous fluids can provide a useful framework to examine the interplay between fluid flow and spatial population genetics.
\end{abstract}

\pacs{}

\maketitle


\section{Introduction}


The transport of living organisms by fluid flows plays an important part in the natural world. Hydrodynamic transport shapes and reorganizes populations across all scales \cite{Tel2005}, mixing populations to uniformity or leading to the formation of spatial structures. For instance, turbulent mixing near the surface of oceans and lakes can cluster phytoplankton blooms into  patchy, fractal-like spatial structures \cite{Abraham1998,Abraham2000} that lead to ecological niches and genetic heterogeneity \cite{dovidio2010,Johnson2006,follows2007}.

Microbial populations expanding into unoccupied territory on agar plates, or range expansions, have been used as a model system to investigate how population spatial structure impacts evolution \cite{Hallatschek2007}. Range expansions develop spatial structure because a thin layer of cells at the population front divide and generate genetically similar daughters who are not pushed very far away before they themselves divide. As a result of this linear population bottleneck at the frontier, the colony loses genetic diversity as the expansion progresses and quickly segregates into large monoclonal sectors that reveal the evolutionary history of the colony in a process often referred to as ``genetic demixing'' \cite{Hallatschek2007}. Simplified stepping stone models with radial inflation   have been used to describe the evolutionary dynamics of this process \cite{Korolev2010}. Microbial range expansions revealed how various evolutionary forces, including selection \cite{Korolev2012a,Gralka2016b,Weinstein2017}, mutualism \cite{Muller2014}, competitive exclusion \cite{Weber2014,McNally2017}, and irreversible mutation \cite{Lavrentovich2016}, impact the dynamics of spatially structured populations.

Microorganism  growing on agar plates cannot be advected as the underlying substrate is a solid, mimicking expansions on land. Although investigated theoretically \cite{pigolotti2013,pigolotti2012a,Perlekar2010,Chotibut2016}, few laboratory systems exist to systematically study the interplay between the transport by fluid flow and spatial population dynamics. In this paper, we introduce a novel experimental system to grow microbial range expansions on the \textit{surface} of a nutrient-rich fluid $10^4$ to $10^5$ times more viscous than water. The extreme viscosity of the liquid substrate enables capillary forces to confine the cells over a macroscopic, quiescent air-liquid interface, and typical settling velocities of isolated cells that leave the surface are less than a cell width per day. This unique system allows us to investigate microbial population morphology and genetic segregation patterns on liquid interfaces.

To our surprise, even in the absence of externally imposed flows \cite{[{See also }]weinsteinThesis_2018}, our experiments revealed that colonies of the budding yeast, \textit{Saccharomyces cerevisiae}, induced strong outwards fluid flows in the surrounding substrate many times larger than the colony's natural expansion velocity. Remarkably, these flows arose from \textit{non-motile} organisms, which do not possess, e. g. the flagellar-induced motilty of bacteria \cite{short2006, mathijssen2018}. In this paper, we show how the induced fluid flow impacts colony morphology and genetic segregation patterns as the viscosity of the underlying substrate varies, and investigate the origin of the induced flow. 

Section \ref{sec:summary_of_three_morphologies} summarizes our most important experimental observations about the morphology and spatial population genetics of expanding yeast colonies on liquid substrates, and identifies three regimes: colonies behave as compact circular colonies, circular colonies with fingers, or many solid-like repelling yeast fragments as the substrate viscosity is varied from high to low. In Section  \ref{sec:piv_flow_around_colony}, we describe our measurements of fluid flows generated near the surface of growing colonies and identify two distinct regimes. Experiments in Section \ref{sec:baroclinic_instability_experiments} argue that the fluid flow is not generated by surface tension gradients (Marangoni flows)\ but is instead generated when yeast metabolism decreases the density of the surrounding fluid, generating buoyant fluid flows via a baroclinic instability due to the pressure and density contours crossing each other at an angle in the vicinity of the colony. Fluid-mechanics simulations calibrated to experiment in Section \ref{sec:hydrodynamic_simulations} provide further evidence that the buoyancy-driven baroclinic instability is the source of the fluid flow, as the simulations can quantitatively predict experimental results. Finally, in Section \ref{sec:model_of_dilation}, we present a simple phenomenological model in the spirit of the lubrication-approximation that combines colony growth, expansion, and thinning to predict the critical metabollically induced radial flow velocity at which colonies break apart. We compare predictions from the model to a phase diagram of yeast colony morphology over time as a function of viscosity. The model displays a conventional Fisher population wave in the absence of flow, but predicts exponential growth of the colony radius in the presence of a flow. When this radial flow is too strong, we find a ``thinning catastrophe'', such that the colony thickness tends to zero and breaks apart. Our work suggests many interesting avenues for future exploration, discussed in Section \ref{discussion}.

\section{Range Expansions on Liquid Substrates
\label{sec:summary_of_three_morphologies}}

To ensure a macroscopic quiescent liquid surface, we performed experiments with fluids $10^{4}-10^5$ times more viscous than water. The viscosity of the fluid is controlled by adding 2-hydroxyethyl cellulose, a long chain polymer, to YPD (yeast extract, peptone, dextrose (glucose)) microbial growth medium; see Appendix \ref{sec:materials_and_methods} for additional experimental details. Characteristic polymer concentrations used in our experiments and corresponding substrate viscosities are given in Table \ref{tab:viscosity_table}. Although the fluid has shear-thinning properties for shear rates $\dot{\gamma} \gtrsim 10^{-1}  \; \mathrm{s^{-1}}$ \cite{boutelier2016}, as discussed in Appendix \ref{rheo}, the flow typical shear rates were of the order of $\dot{\gamma} = u/H \sim 10^{-6} - 10^{-5}  \; \text{s}^{-1}$, where $u$ is the characteristic surface flow velocity and $H$ is the substrate fluid height, such that non-Newtonian effects were negligible in our experiment. In contrast with plates filled with hard agar, which form a gel substrate with a shear modulus, cellulose polymers do not form a three-dimensional mesh, allowing the growth medium to flow.

\begin{table}[h]\centering
\ra{1}
\begin{ruledtabular}
\begin{tabular}{@{}cc@{}}
Polymer \% (w/v)        & $\eta$ (Pa$\cdot$s)\\
\hline
2.0                     & 54 $\pm$ 8\\
2.2                     & 86 $\pm$ 13\\
2.4                     & 140 $\pm$ 20\\ 
2.6                     & 300 $\pm$ 45\\
2.8                     & 450 $\pm$ 70\\
3.0                     & 600 $\pm$ 90\\
\end{tabular}
\end{ruledtabular}
\caption{Newtonian approximation to the liquid substrate's viscosity at a shear rate of $\dot{\gamma} \sim 10^{-4}\; \text{s}^{-1}$(Appendix \ref{rheo}) at various concentrations 24 hours after mixing it with 2-hydroxyethyl cellulose.
\label{tab:viscosity_table}
}
\end{table}

\begin{figure*}
\centering
\includegraphics[width=17.8cm]{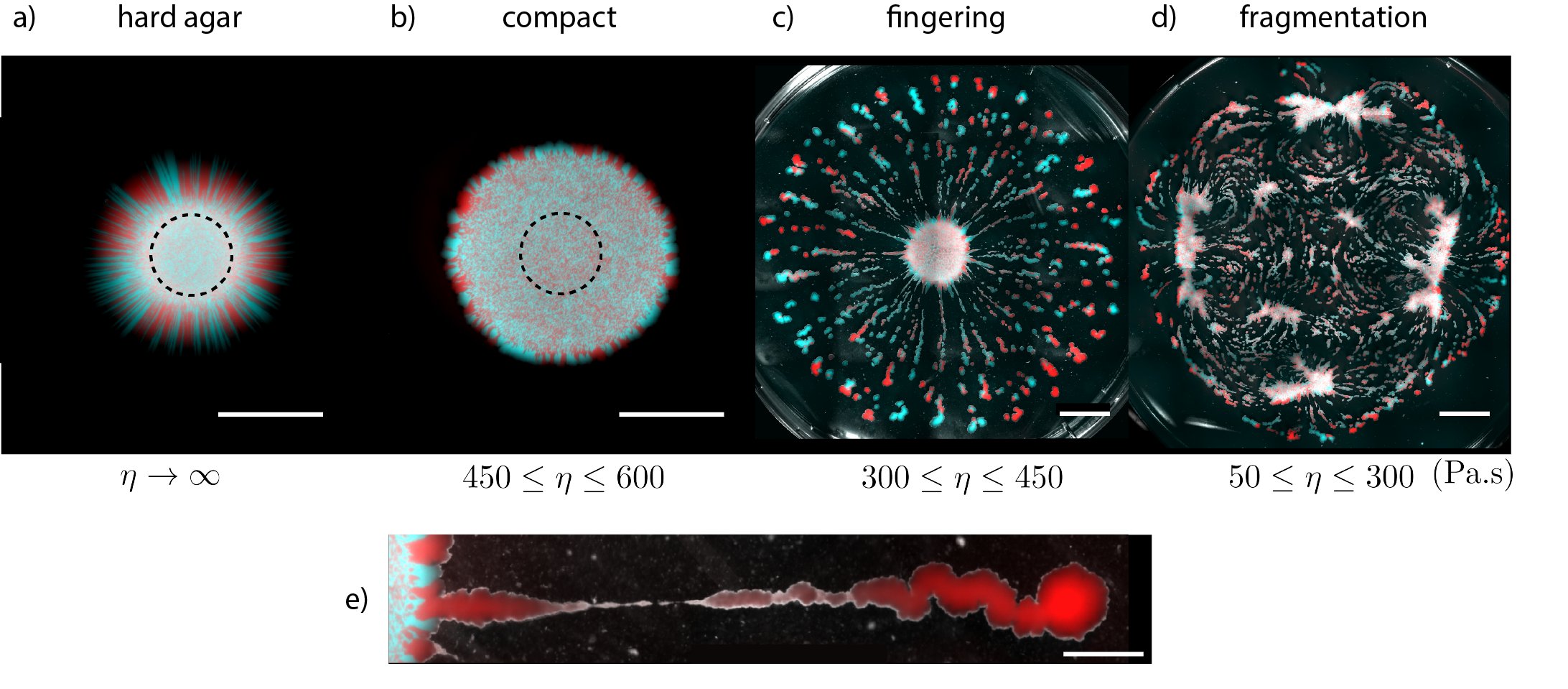}
\caption{Selected yeast colony morphologies on a) a hard agar plate after 72h of growth, and on the surface of the viscous substrate with decreasing viscosities: b) for $\eta = 600\pm90$ Pa$\cdot$s after 72h of growth, c) $\eta = 450 \pm70$ Pa$\cdot$s after 84h of growth, d) $\eta = 300\pm45$ Pa$\cdot$s after 36h of growth, and e) magnification of a single representative finger from regime c). Qualitatively similar morphologies were observed in the range of viscositites indicated in b)-d). The figure shows merged brightfield and fluorescent images. White: transmitted brightfield, red: YFP strains, and cyan: mCherry strains. The scale bars in a) and b) correspond to 5 mm, the scale bars in c) and d) to 10 mm, and to 2 mm in e). 
\label{examples}}
\end{figure*}

Our large substrate viscosity prevented thermal gradients in the environment from driving undesired convection under our laboratory conditions; no substrate fluid motion was observed due to stray thermal gradients in the absence of a colony growing on the surface. After deposition on the substrate, droplets containing yeast cells spread unformely, allowing a dilute concentration of cells to be held at the air-liquid interface by capillary forces. The cells rapidly aggregate due to attractive forces: capillary forces at the interface \cite{Vella2005} for large distances, and Van der Waals forces between the cells for short distances, in a process resembling spinodal decomposition or nucleation and growth \cite{Langer1971}; see supplementary Figure \ \ref{fig:yeast_spinodal_decomposition} and supplemental movie 0. Capillary forces were large enough to keep the cells on the surface of the fluid despite their slightly higher density than the media, allowing the colony to grow at the air-liquid interface over the typical several days time scale of our experiments. The large substrate viscosity also leads to extremely slow sedimentation velocities of any small clumps of cells that break through the surface.

We followed the segregation of two \textit{S. cerevisiae} strains, genetically identical except for constitutively expressing different fluorescent proteins. The experiments were initiated by depositing cells in a $2$ $\mu$L droplet of saturated overnight culture at the center of a $94$ mm diameter circular Petri dish filled with 40 mL of our viscous medium. The resulting colony expansion was then monitored over 5 days with a stereoscope (Appendix \ref{sec:materials_and_methods}). Shortly after cell growth and division begin, the microorganisms exhibit dramatically different growth dynamics relative to the well-studied hard agar plates, and exhibit a rich variety of morphologies depending on the media viscosity. We systematically varied the polymer concentration in the medium, allowing us to investigate the microbial population behavior over a range of dynamic viscosities $\eta$ from $54\pm8$ Pa$\cdot$s to $600\pm90$ Pa$\cdot$s (corresponding to 2\% to 3\% w/v polymer, Table \ref{tab:viscosity_table}). Figure \ref{examples} shows examples of yeast colonies after $72$ hours of growth on a hard agar gel plate, compared to growth on liquid substrates for three different viscosities.

At the highest viscosity studied, $\eta = 600\pm90$ Pa$\cdot$s, the yeast cells formed a single, compact, circular colony which expands radially over time (see supplemental movie 1).
However, unlike colonies on solid media where genetic drift dominates very close to the original frontier of the inoculation \cite{Hallatschek2007}, colonies on the substrate had a stretched central region with genetic diversity (two colors were mixed together); demixing only occurred at a much larger colony radius, as displayed in Figs \ref{examples}(a) and \ref{examples}(b) where the size of the initial inoculum is shown as a black dashed circle.
Genetic domain walls with neutral strains impinge at right angles to a colony's front and are driven by interfacial undulations \cite{Hallatschek2007}.
Yeast cells grown on the viscous liquid presented much rougher colony fronts than on hard agar plates, leading to more irregular domain walls after the onset of genetic demixing.
As the viscosity decreased to $\eta \approx 450$ Pa$\cdot$s, the initially circular colony formed numerous smaller microbial assemblies at its periphery on the media's surface. The front of the originally circular colony became unstable and finger-like structures formed within the first $24$h of growth; a large fingering colony spanning an entire Petri dish after 84 hours of growth can be seen in Figure \ref{examples}c) and supplemental movie 2; a high-magnification picture of a finger is shown in the bottom panel of Fig. \ref{examples}c). These fingers form after demixing has occurred, typically leading to monoclonal aggregates that grow and break up into small clusters, somewhat reminiscent of a Plateau-Rayleigh instability \cite{rayleigh1878, Eggers1997a}. However, our system is complicated by active cell divisions and a colony-generated radial velocity field (see Sec. \ref{sec:piv_flow_around_colony}).
Below $\eta = 300\pm45$ Pa$\cdot$s, the initial colony fractured into irregular pieces within the first 12 hours of expansion, behaving as if they had a shear modulus on our experimental time scales, and formed highly fragmented colonies as seen in Figure \ref{examples}d) and supplementary movie 3. Colonies in this regime break apart before genetic demixing occurred, resulting in genetically diverse growing fragments. The regularly interspersed fragments repel each other as they continue to grow, suggesting the existence of an underlying repelling flow. At the lowest studied viscosity, $\eta = 54\pm8$ Pa$\cdot$s, these clusters of yeast cells propelled themselves across an entire Petri dish within 36 hours, dispersing more than one order of magnitude faster than the same yeast strains growing on 2\% hard agar plates (see Fig. \ref{fronts} for the radial growth of our strains on agar and liquid substrates over time).

\section{Colony-Generated Flow \label{sec:piv_flow_around_colony}}

\begin{figure}
\includegraphics[width=8.6cm]{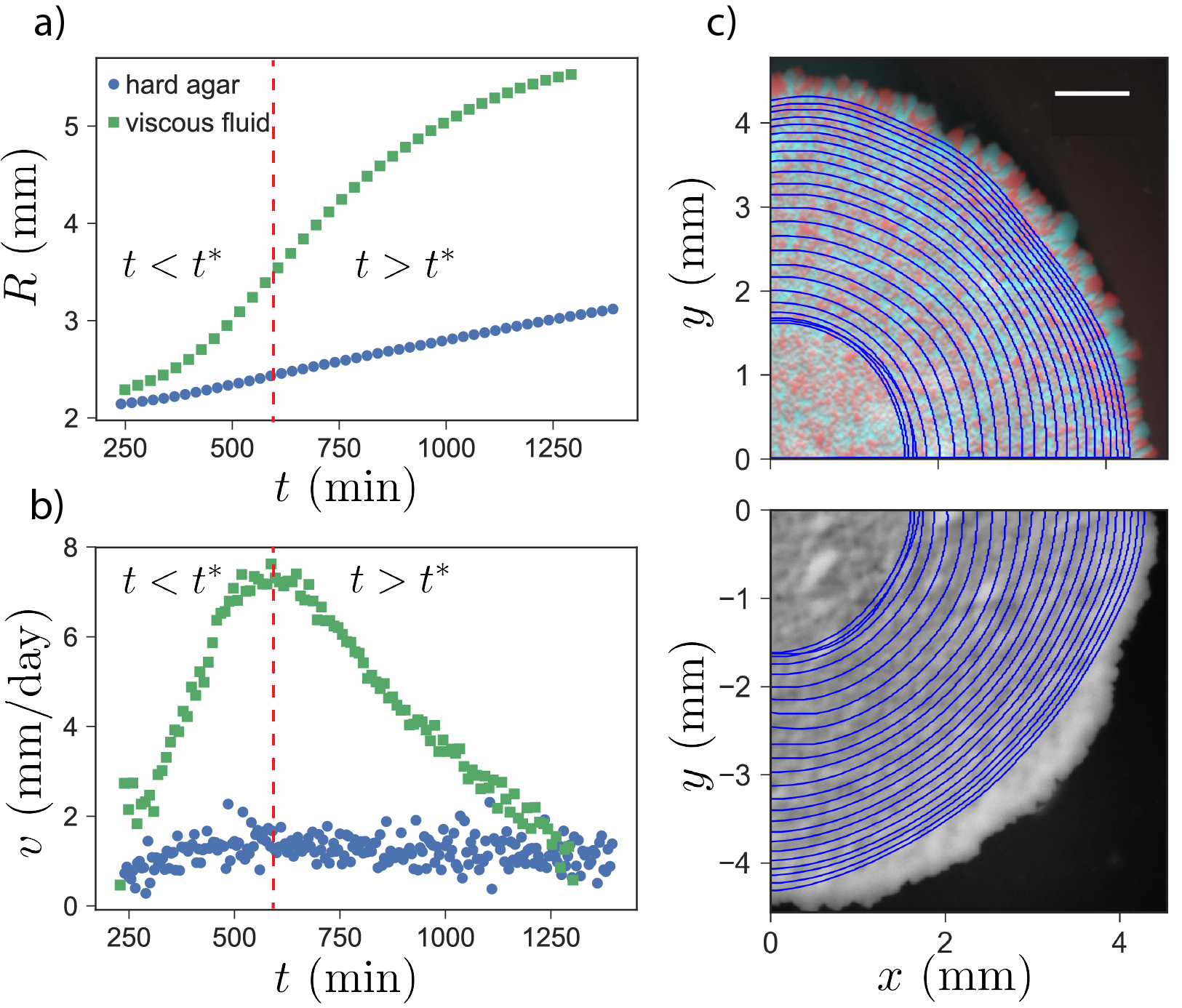}
\caption{a) Azimuthally averaged yeast colony radius $R(t)$ during the first 24h of growth on hard agar, blue circles, and on a liquid substrate with viscosity $\eta = 600\pm 90$ Pa$\cdot$s, green squares. b) The corresponding colony front velocity extracted from $R(t)$, the colony exhibits two growth regimes on the liquid substrate: a superlinear regime for $<t^\ast$ and a slowly decaying phase for $t>t^\ast$. We found that the colony front velocity approaches $v(t) = 0.5 \pm 0.05$ mm/day at long times ($t \gg\ t^\ast$) which is less than the velocity of yeast colonies growing on 2\% hard agar plates. c) Consecutive front spatial positions at equal 40 min intervals during the first 24h of growth on liquid substrate with the same viscosity as in a) and b), overlayed on top of a fluorescent image, top, and on brightfield image of the colony, bottom. Note that genetic demixing begins at the edge of the colony after the front has slowed down. The scale bar corresponds to 1 mm.
\label{fronts}}
\end{figure}

In this Section, we focus, for simplicity, on the high viscosity regime $450 \lesssim \eta \lesssim 600$ Pa$\cdot$s where yeast cells form a single, approximately circular colony to investigate the coupling between its growth and the three-dimensional fluid flows generated in its vicinity. We imaged yeast colonies growing during the first 48 hours after inoculation and extracted in parallel the fluid velocity near the substrate's surface with particle image velocimetry (PIV). The fluid was seeded with a dilute concentration of $10-20 \ \mathrm{\mu m}$ fluorescent, neutrally buoyant polyethylene beads, and horizontal slices of the flow were followed at the desired height by varying the focal plane at which the beads motion was tracked; see more details in Appendix \ref{sec:materials_and_methods}. Figure \ref{fronts} displays the expanding colony average radius $R(t)$, velocity $v(t)$ and two-dimensional front profile over time extracted from brightfield images.

\begin{figure*}
\includegraphics[width=17.8cm]{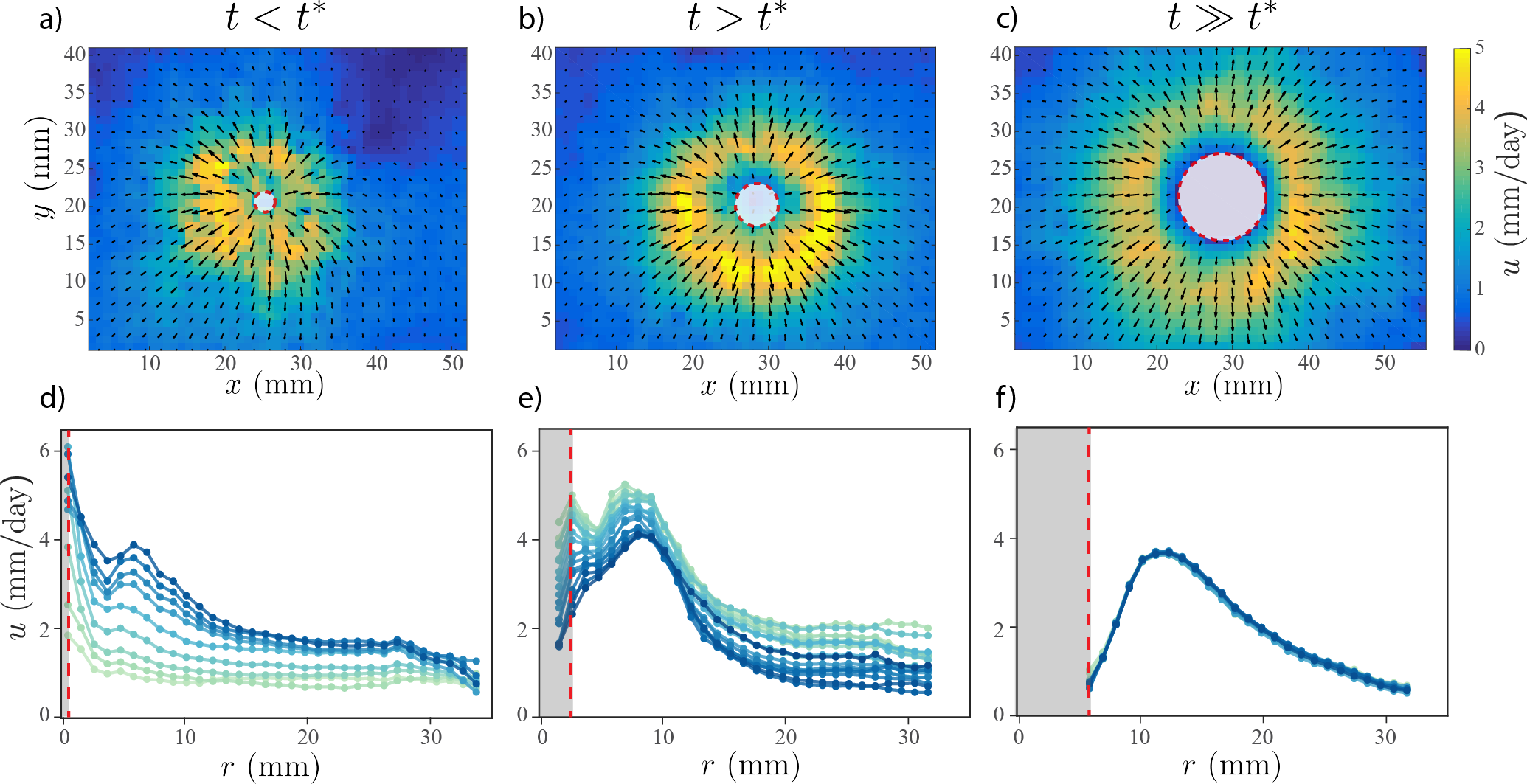}
\caption{
Experimental flow field at the viscous substrate's surface over the first 48 hours of a compact yeast colony growth for a substrate viscosity of $\eta = 600\pm90$ Pa$\cdot$s. The central gray region delineated by red dashed lines indicates the growing colony's radius positions masking the fluorescent beads; we could not directly measure the velocity below the colony with this experimental setup. The velocity field in the vicinity of the colony was averaged over 3h for $t<t^\ast$ (a), $t>t^\ast$ (b), and $t \gg t^\ast$ (c). The colormap represents the flow velocity amplitude. The azimuthal average of the velocity radial profile is plotted every 10 min for $t<t^\ast$ (d), and every 20 min for $t>t^\ast$ (e) and $t \gg t^\ast$ (f). Lighter lines correspond to earlier times.
\label{piv}}
\end{figure*}

In contrast to yeast cells growing on hard agar plates which expand with approximately constant radial velocity \cite{Hallatschek2007,Korolev2012a,Muller2014,Gralka2016b}, two distinct growth regimes separated by a characteristic time $t^\ast \approx 600$ min can be identified on liquid substrates. At early times for $t<t^\ast$, the colony radius expands superlinearly with time and reaches a maximum horizontal growth velocity of $v \simeq 7.5 \pm 0.8 \ \mathrm{mm/day}$, while for $t>t^\ast$ the expansion rate gradually slows down to $v \simeq 0.5 \pm 0.05 \ \mathrm{mm/day}$ over the rest of the experiment as shown in Fig. \ref{fronts}b).
This first, approximately exponential, growth regime when $t<t^\ast$ suggests that cells dividing throughout the entire colony contribute to its surface area expansion, in contrast to growth on hard agar where only cells dividing near the front of the colony contribute to its expansion \cite{Hallatschek2007}. A comparison of the expansion rate of the colony with the spatial distribution of the strains reveals that genetically demixed sectors appear only after the front propagation slowed down to $v \lesssim 2\ \mathrm{mm/day}$, as shown in Figure \ref{fronts}c), when only those regions exhibiting demixing at the edge of the colony are growing (see supplemental movie 4).

PIV measurements carried out in the same experiment near the surface of the fluid revealed an outward radial flow centered around the colony which began soon after the first cell divisions occurred; two-dimensional snapshots of the velocity field are displayed in Figs. \ref{piv}a), \ref{piv}b) and \ref{piv}c) for $t<t^\ast$, $t>t^\ast$ and $t\gg t^\ast$ respectively, while Figures \ref{piv}d), e) and f) display the evolution of the azimuthal average of the velocity field $u_r(r,t) \equiv u(r,t)$ over time. The flow is radially symmetric, reflecting the circular colony shape at high viscosity, and its overall magnitude increases within 24 hours after inoculation.
Two distinct regimes can be identified. At early times, for $t<t^\ast$, the radial velocity profile exhibits a maximum near the edge of the growing colony, whose value increases in time, peaking at $u = 6 \pm 0.8 \ \mathrm{mm/day}$ for $t \simeq 560$ min after inoculation, and rapidly decreases away from the colony. The similar values and variation exhibited by the colony front propagation velocity $v(t)$ for $t < t^\ast$, and displayed on Fig. \ref{fronts}(b), suggest that the fluid is radially pushed outwards by the exponentially expanding colony during this time period.

However, as the expansion slows down after $t^\ast$, a secondary peak with a smaller amplitude can be observed in Fig. \ref{piv}d) and e). Within 48 hours it approaches a time-independent velocity $u= 4 \pm 0.5 \ \mathrm{mm/day}$, shown in Figs. \ref{piv}c), at about 1.5 colony radii away from the colony center despite the fact that the colony expansion velocity had slowed to $v(t) \lesssim 0.5 \pm 0.05 \ \mathrm{mm/day}$. These observations suggest that the expanding edge of the colony pushing the surrounding fluid is not the unique origin of the observed flow and another mechanism is generating the flow in the surrounding media for $t\gg t^\ast$, an idea we pursue in the next Section.

\section{Baroclinic Instability \label{sec:baroclinic_instability_experiments}}

Plates filled with viscous media and monitored over 24 hours under conditions identical to our experiments showed no evidence of flow in the absence of growing yeast cells, suggesting that the colony metabolism is responsible for the flow observed at $t>t^\ast$. A wide variety of microbial organisms exploit Marangoni flows \cite{Scriven1960} to facilitate their horizontal displacement across liquid interfaces by locally reducing the surface tension \cite{angelini2009,Yang2017,Kearns2010}. Yeast cells secrete a wide variety of molecules in their vicinity, including ethanol and pheromones, which could potentially lower the substrate surface tension in the colony surrounding. Surfactant-releasing particles, such as camphor boats, can lead to the formation of mutually-repelling assemblies \cite{Soh2008} similar, for example, to the fragmented yeast aggregates we observe under the experimental conditions shown in Fig. \ref{examples}d) and \ref{pangaea}. On the other hand, the yeast cell metabolism could also generate large enough gradients in the surrounding fluid's temperature or solute concentration, to produce local differences in density and drive buoyant flows in the presence of a gravitational field \cite{Turner1973}.
However, as shown in the work of Benoit et al.\ \cite{Benoit2008}, temperature gradients can be ruled out because heat diffuses over 200 times faster than small-molecule solutes (such as glucose) in water, minimizing resulting density gradients, and because the coefficient of thermal expansion is so much smaller than the coefficient of solute expansion; large temperature differences (several degrees Celsius) would be required to create the same density difference as a small change in solute concentration (see Appendix C for details).

\begin{figure}
\includegraphics[width=8.6cm]{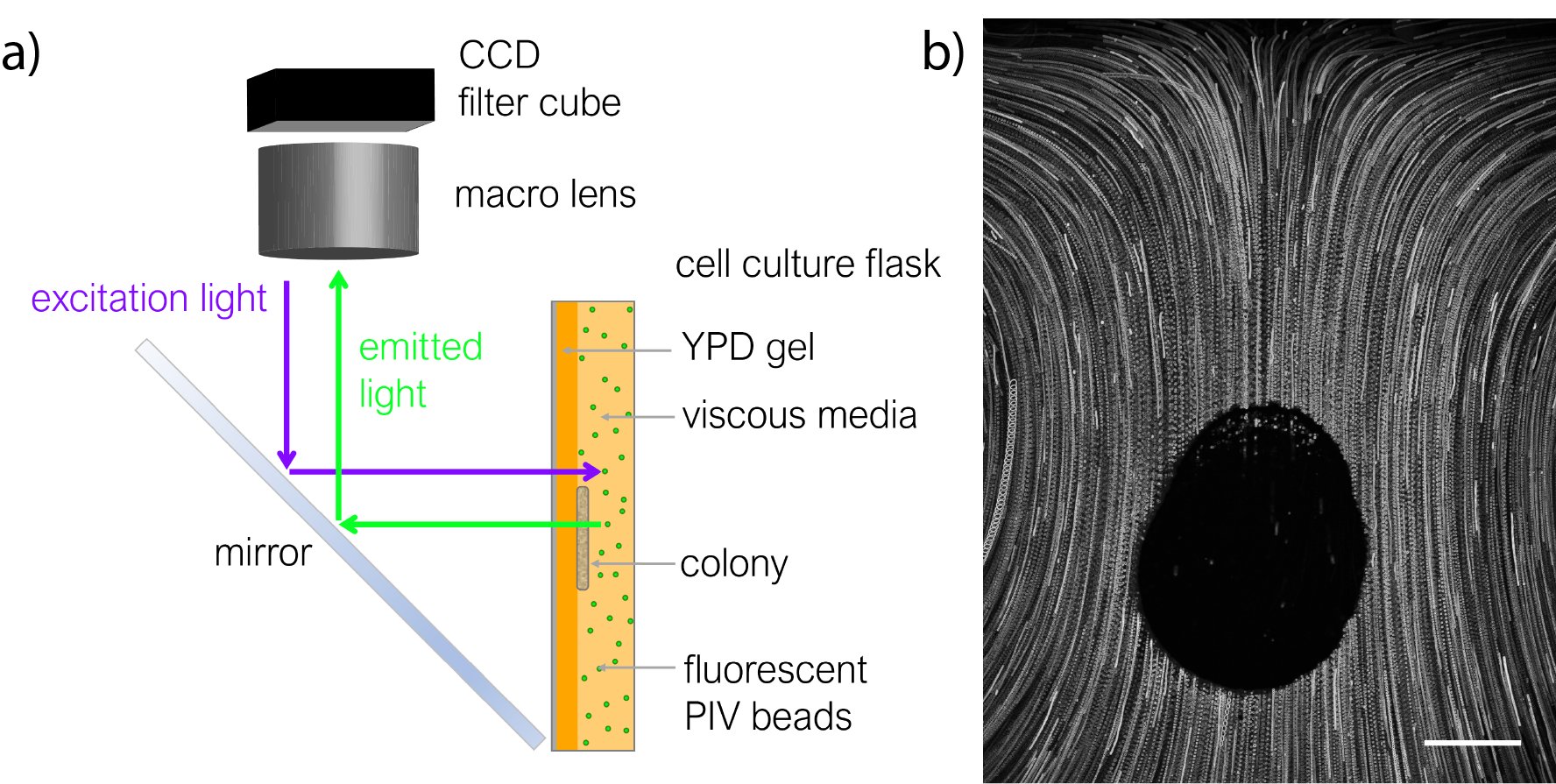}
\caption{a) Experimental setup for a yeast colony anchored on the side wall of a sealed chamber filled with the viscous liquid; no liquid-air interfaces were present, removing the possibility of Marangoni flows. Gravity points downward, and the fluid was seeded with fluorescent PIV beads to track fluid motion. b) Fluid flow streamlines over the yeast colony (the dark circular patch) during a time interval of $\Delta t \approx 6$h; obtained via maximum intensity projection. The scale bar corresponds to 5 mm.
\label{scheme}}
\end{figure}

In order to discriminate between these different sources of flow, we conducted a series of experiments where we anchored the colonies on a thin layer of agar to the top, bottom, and sides of sealed chambers filled with our viscous media (see Fig. \ref{fig:buoyant_flow_directions_and_plume} for details). We found that colonies created fluid flows similar in magnitude to experiments when the air-liquid interface was present, and regardless of their position in the chamber (even when placed at the top of a sealed chamber). The induced fluid flows always opposed the direction of gravity, such as the one shown in Fig. \ref{scheme}a), where a colony attached to a vertical wall entirely immersed in the liquid media created an upwards flow over its surface; one large vortex on each side of the colony is partially visible in Fig.\ \ref{scheme}b). Although these experiments did not rule out the possibility that surface tension gradients affect the flow when a free interface is present, they revealed that buoyant forces are primarily driving the observed flows.

The flow was systematically opposing the direction of gravity regardless of the position of the colony in the sealed chambers, suggesting that the cells' metabolism altered the density of the substrate by \textit{depleting} nutrients in the surrounding fluid, for instance by taking biomass from the solute to create progeny or by converting denser solute molecules into lighter ones (e.g. fermentation converts glucose to ethanol and carbon dioxide which are both less dense than glucose in water). In fact, similar behavior has been observed from \textit{E. coli} growing in sealed chambers filled with liquid media \cite{Benoit2008}. Measuring the initial and final density of the medium after a yeast culture grew to saturation in YPD showed a decrease in density $\Delta \rho = -0.0090 \pm 0.0005  \ \mathrm{g/mL}$, where the $\pm$ corresponds to the range of density differences we measured (see Appendix \ref{sec:materials_and_methods} for additional details), confirming that proliferating yeast cells reduce the density of the surrounding media.

However, in contrast to microbes growing at the bottom of liquid-filled sealed container that can induce a classical Rayleigh-Taylor instability \cite{Sharp1984,Guyon,Benoit2008}, where the less dense fluid near the colony rises, the cells in our experiments grew on the surface of a liquid-air interface and could not generate flow with this particular instability. Instead, the yeast produces a localized pocket of less dense fluid \textit{on top} of a more dense fluid. In this configuration, the resulting density contours' misalignment with the hydrostatic pressure horizontal isobars leads to a thresholdless baroclinic instability. This type of instability, common in stratified fluids, generates vorticity and can be observed in atmospheric and oceanic flows \cite{Turner1973,Guyon,marshall2008}.

The origin of the instability can be understood starting with the Navier-Stokes equations for the substrate fluid:
\begin{equation}
\frac{\partial \mathbf{u}}{\partial t} + (\mathbf{u} \cdot \boldsymbol{\nabla})\mathbf{u} =
- \frac{1}{\rho}\boldsymbol{\nabla}p + \nu \nabla^2 \mathbf{u} + \mathbf{g},
\label{NS}
\end{equation}
where $\mathbf{u}$ is the fluid velocity, $\rho$ the fluid density, $p$ the pressure, $\nu=\eta / \rho$ the kinematic viscosity of the liquid medium, and $\mathbf{g} = -g\mathbf{\hat{z}}$ the gravitational force. Upon taking the curl of the fluid velocity $\mathbf{u}$ we obtain the vorticity
$\boldsymbol{\omega}=\boldsymbol{\nabla} \times \mathbf{u}$ and find:
\begin{equation}
\frac{\partial \boldsymbol{\omega}}{\partial t} + \left(\mathbf{u} \cdot \boldsymbol{\nabla}\right)
\boldsymbol{\omega} =
\left(\boldsymbol{\omega} \cdot \boldsymbol{\nabla}\right)\mathbf{u}
+ \frac{1}{\rho^2}\left(\boldsymbol{\nabla}\rho \times \boldsymbol{\nabla}p\right) 
+ \nu \nabla^2 \boldsymbol{\omega}.
\label{eq:vorticity_equation}
\end{equation}
In the limit of small flow velocities, second order terms in $\boldsymbol{\omega}$ and $\mathbf{u}$ (vorticity advection and vortex stretching) can be neglected, and Eq. \eqref{eq:vorticity_equation} simplifies to:
\begin{equation}
\frac{\partial \boldsymbol{\omega}}{\partial t}
\approx \frac{1}{\rho^2}\left(\boldsymbol{\nabla} \rho \times \boldsymbol{\nabla} p \right) + \nu \nabla^2 \boldsymbol{\omega}.
\label{baro_eq}
\end{equation}
The viscous term $\nu \nabla^2 \boldsymbol{\omega}$ simply redistributes the vorticity in the bulk fluid. However, the term $\frac{1}{\rho^2}\left(\boldsymbol{\nabla} \rho \times \boldsymbol{\nabla} p \right)$, often called the ``baroclinicity'' \cite{marshall2008}, generates vorticity whenever the contours of constant density $\rho$ and pressure $p$ cross at a finite angle.


\section{Hydrodynamic Simulations \label{sec:hydrodynamic_simulations}}


\begin{table*}\centering
\caption{
Model parameters and their experimentally measured values, where appropriate. For additional details, see the Appendix \ref{sec:calibrating_simulation}. Unless otherwise indicated, the error bars correspond to the standard deviation.
\label{tab:model_parameters}}
\begin{ruledtabular}
\begin{tabular}{lllp{3.5in}}
Parameter & Value & Units & Description\\
\hline
$\nu $ & $100-1000$  &$\mathrm{cm^2/s}$ & Kinematic viscosity; varies with polymer concentration\\

$D$ & $2.4\pm 0.3 \times 10^{-6}$  & $\mathrm{cm^2/s}$ & Diffusion coefficient of small nutrient molecules \\

$\rho_1$  & $1.015 \pm 0.005$  &$\mathrm{g/mL}$ & Density of the viscous substrate with nutrients\\

$\beta c_1$ & $0.009\pm 0.001$ &  $\mathrm{None}$ & Product of the expansion coefficient $\beta$ and $c_1$\\

$a c_1$ & $5 \pm 2$ &$\mathrm{pg/(\mu m^2 h)}$ & Product of the mass flux into the yeast colony $a$ and $c_1$\\

$H$ & $1-10 $ &$\mathrm{mm}$ & Fluid height in the Petri dish ($h\approx 7$ mm for 40 mL) \\

$r_\text{petri}$ & $43 \pm 1$ & $\mathrm{mm}$ & Radius of the petri dish\\

$|\mathbf{g}|$ & $9.81$  &$\mathrm{m/s^2}$ & Gravitational acceleration\\

$R$ & $1 - 8$ & $\mathrm{mm}$ & Average radius of a yeast colony during an experiment\\ 
                             
$\ell \equiv \rho_0 \beta D/a$ & $1.6 \pm 0.8$ & $\mathrm{mm}$ & Characteristic nutrient depletion length in the fluid. \\
\end{tabular}
\end{ruledtabular}
\end{table*}

\subsection{Origin of the Baroclinic Instability \label{sec:sim_origin_of_baro}}


To better understand how yeast colonies living at a liquid interface can trigger a baroclinic instability, we first assume a fluid at rest and numerically investigate how baroclinicity is created as the cells deplete the surrounding nutrient field by examining the resulting density and pressure contours. We assume the fluid has a density $\rho$ which depends on the local concentration field $c(\mathbf r,t)$ of a diffusing nutrient solute such as glucose. The solute concentration is depleted near the metabolizing yeast cells such that the mass density of the fluid, given by
\begin{equation}
\rho(\mathbf{r},t)=\rho_0 + \delta \rho(\mathbf{r},t) = \rho_0\left[ 1+\beta c(\mathbf{r},t) \right] ,
\end{equation}
locally decreases, where $\rho_0$ is the fluid density without nutrient solute, $\beta = \frac{1}{\rho_0} \left( \frac{\partial \rho}{\partial c} \right) $ is the solute expansion coefficient, and $ \delta \rho(\mathbf{r},t) = \rho_0\beta c(\mathbf{r},t)$ gives the local increase in density due to the presence of nutrients \cite{Benoit2008}. Let $c_1$ be the initial reference nutrient concentration before any metabolic depletion occurs, such that, close to the metabolizing colony, there is a reduction in $\rho(\mathbf{r},t)$ and $c(\mathbf{r},t)<c_1$. In the absence of a flow, the momentum equation \eqref{NS} simplifies to a hydrostatic pressure balance coupled to nutrient diffusion in the substrate fluid and becomes:
\begin{eqnarray}
- \boldsymbol{\nabla}p + \rho \mathbf{g} &=&0 \label{hydrostat} \\
\frac{\partial c}{\partial t} &=&  D \nabla^2 c , \label{nut_diff}
\end{eqnarray}
where $D$ is the diffusion constant of the nutrient solute molecules.

\begin{figure}
\includegraphics[width=8.6cm]{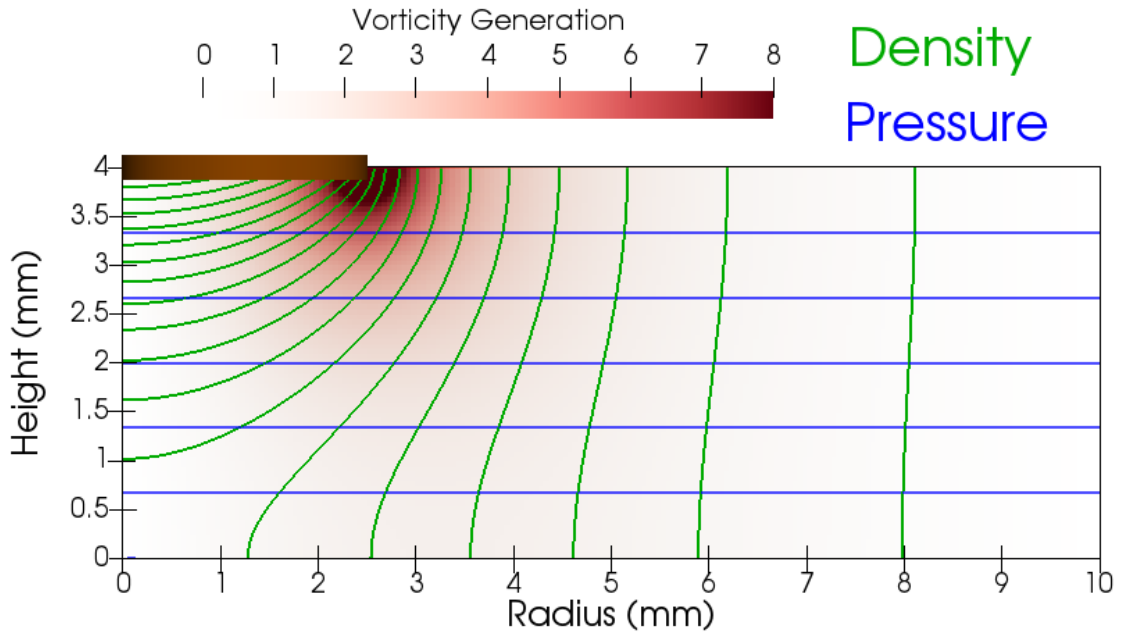}
\caption{
Baroclinic vorticity generation rate $\partial \boldsymbol{\omega} / \partial t \approx \frac{1}{\rho^2}\left(\boldsymbol{\nabla} \rho \times \boldsymbol{\nabla} p \right)$ normal to a radial cross-section before flow is initiated by a yeast colony fixed at the surface of the viscous fluid in a radially symmetric Petri dish. The colony position is indicated by the thick orange line, the pressure isobars in blue, and the density contours in green. The isobars are near-horizontal due to small density differences originating from nutrient depletion (in this simulation, $\Delta \rho_\text{max}\sim -0.003 \ \mathrm{g/mL}$). Whenever the pressure and density contours cross at an angle, vorticity is generated via the baroclinic term in Eq. \eqref{eq:vorticity_equation}.
\label{baro}}
\end{figure}

We account for the colony nutrient absorption by imposing a nutrient mass flux normal to the colony's surface $\mathbf{j_{\mathrm{col}}}= a c \mathbf{\hat{n}}$, where $a$ is the mass flux rate into the colony per unit nutrient concentration and $\mathbf{\hat{n}}$ is the unit normal vector to the interface, such that larger nutrient concentrations lead to a larger nutrient absorption rate \cite{Lavrentovich2013}. In contrast, no-nutrient-flux boundary conditions are applied elsewhere, on the walls of the domain away from the colony, $D\boldsymbol{\nabla}c \cdot \mathbf{\hat{n}}=0 $.
The mass flux due to transport and diffusion in the bulk fluid is given by $\mathbf{j_{\mathrm{fluid}}} =  \rho_0\beta \left(\mathbf{u}c - D\boldsymbol{\nabla} c \right)$. We assume that $\mathbf{u}=0$ for now, and upon applying continuity on the solute flux across the colony boundary $\left(\mathbf{j}_\text{colony}=\mathbf{j}_\text{fluid}\right)\big|_\text{colony}$, the boundary condition can be rewritten as:
\begin{equation}
\left(\boldsymbol{\nabla} c \cdot \mathbf{\hat{n}}\right)\bigg|_\text{colony} = \frac{c}{\ell} \bigg|_\text{colony}
\label{yeast_bc}
\end{equation}
where $\ell = \rho_0 \beta D / a =1.6 \pm 0.8 \ \mathrm{mm}$ acts as a characteristic nutrient depletion length in the fluid that captures the interplay between nutrient diffusion and absorption by the bottom of the yeast colony. Here, $\ell$\ is different than the nutrient screening length \textit{inside} the yeast colony \cite{Lavrentovich2013}, as discussed in Appendix \ref{sec:calibrating_simulation}. Note that our yeast cells do not absorb the concentration field fast enough to warrant setting $c=0$ at the interface between the colony and the fluid substrate as indicated by the dimensionless numbers discussed in Appendix \ref{sec:nondimensionalizing_flow}.

The actual colony expansion is neglected for simplicity, so we consider a colony of fixed radius $R$ at the surface of the viscous fluid, in a radially symmetric Petri dish as shown in Figure \ref{baro}; the yeast colony is represented by the thick orange line. We use OpenFOAM 5.0 \cite{openfoam} to simulate equations \eqref{hydrostat}-\eqref{yeast_bc} using the program \texttt{diffusionPressureFoam} \cite{stokesBuoyantSoluteFoam} and the measured parameters from Table \ref{tab:model_parameters}; additional details about the numerical scheme appear in Appendix \ref{sec:simulation_methods}. Figure \ref{baro} displays the resulting density contours and isobars. Once the cells start absorbing nutrient mass from the fluid, a curved density gradient that conforms to the finite size of the colony is created in its vicinity; supplementary Fig. \ref{fig:simulation_cfield_and_ufield} shows an example of a corresponding simulated concentration field. The pressure contours, on the other hand, remain nearly horizontal over the entire domain as the density differences due to solute depletion are so small. The finite crossing angle of the pressure and density contours leads to vorticity generation via the baroclinic term $\frac{1}{\rho^2}\left(\boldsymbol{\nabla}\rho \times \boldsymbol{\nabla}p\right)$ in Eq. \eqref{baro_eq} below the edge of the yeast colony, where the gradient of density is large and nearly perpendicular to the pressure gradient. As long as the yeast cells deplete the surrounding nutrients, the created density difference will generate vorticity via this thresholdless baroclinic instability.

\subsection{Comparison with Experiment \label{sec:sim_comparison_with_experiment}}

\begin{figure}
\includegraphics[width=8.7cm]{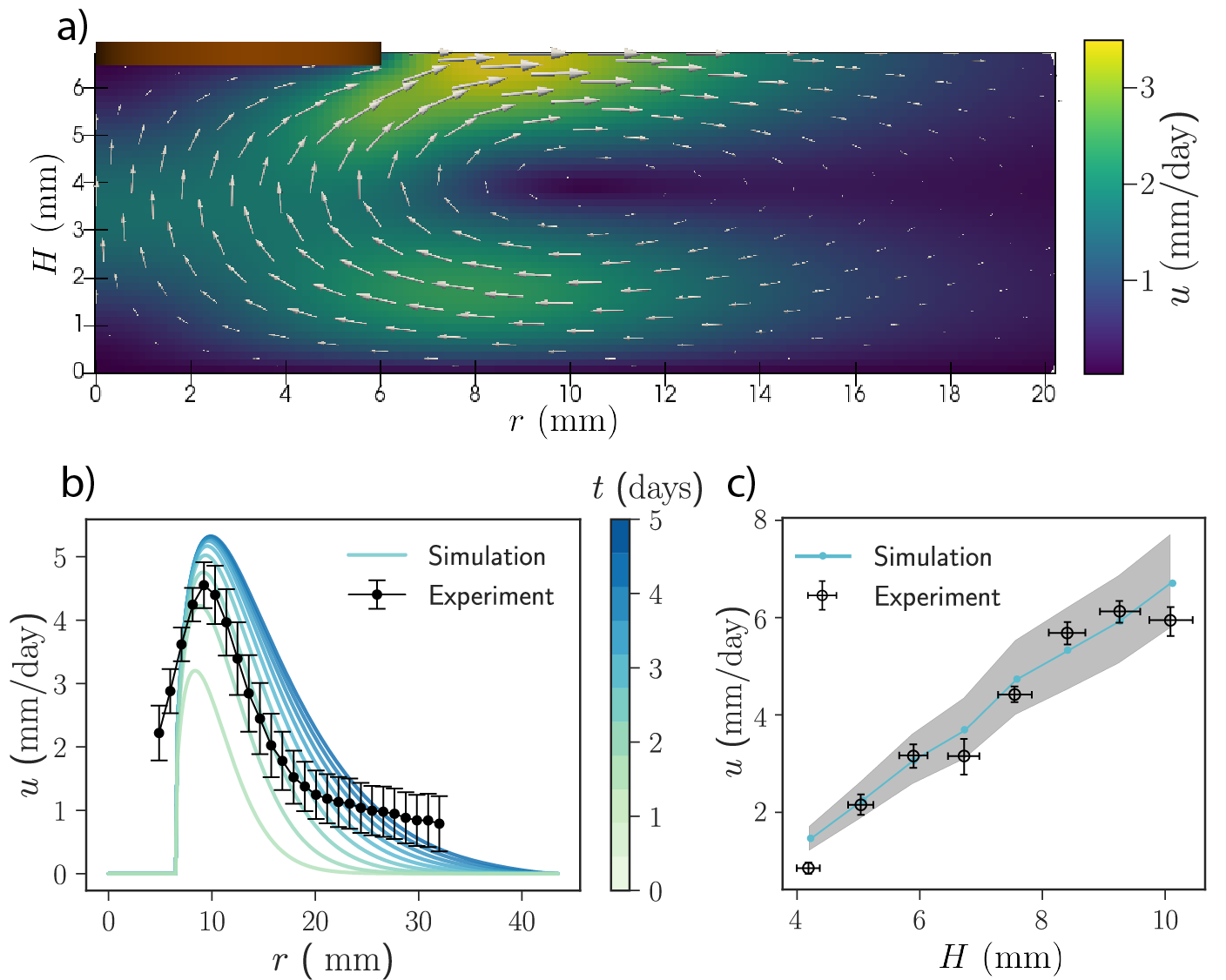}
\caption{
a) Snapshot of the simulated flow field below the yeast colony (brown bar) after flow is initiated, for $t \gg t^\ast$. The simulated flow field qualitatively matches our experiments with a vortex ring produced around the colony. b) Azimuthal average of the numerical flow field using the measured parameters in Table \ref{tab:model_parameters} plotted every 12 hours at the substrate fluid surface.  Black circles, experimental flow
radial profile measured for similar flow parameters after 24 hours of growth and an initial $\eta=600 \pm 90 \ \mathrm{Pa\cdot s}$. c) Simulated and experimental peak radial velocity, determined from PIV measurements, as a function of fluid height below the colony. The blue line with circles corresponds to the simulated values using the parameters in Table \ref{tab:model_parameters}, the black shaded region is the standard deviation of the simulated points, and the black circles corresponds to experimental data.
\label{simu_flow}}
\end{figure}

We now determine the flow produced by the baroclinic instability in the liquid substrate by simulating the hydrodynamic flow equations, and compare our simulations with the experimental flow velocities.
The diffusing solute field is now coupled with the incompressible Navier-Stokes equations, and we now must solve the full set of equations,
\begin{eqnarray}
\frac{\partial c}{\partial t} + \mathbf{u}\cdot \boldsymbol{\nabla} c &=& D \nabla^2 c ,\\
\frac{\partial \mathbf{u}}{\partial t} + \left( \mathbf{u} \cdot \boldsymbol{\nabla} \right)\mathbf{u} &=& - \frac{1}{\rho} \boldsymbol{\nabla}p +\nu \nabla^2 \mathbf{u} + \mathbf{g},\\
\boldsymbol{\nabla} \cdot \mathbf{u} &=& 0.
\end{eqnarray}
In the limit of small local density variations, $\delta \rho(\mathbf{r}) / \rho_0 \ll 1$, we can apply the Boussinesq approximation \cite{Turner1973,Benoit2008}, such that equation \eqref{NS} becomes:
\begin{equation}
\frac{\partial \mathbf{u}}{\partial t} + \left( \mathbf{u} \cdot \boldsymbol{\nabla} \right)\mathbf{u} =- \frac{1}{\rho_0} \boldsymbol{\nabla}p^\prime + \nu \nabla^2 \mathbf{u} + \beta c(\mathbf{r},t) \mathbf{g},
\label{boussi}
\end{equation}
where the pressure $p^\prime=p - \rho_0 gz$ is the pressure measured relative to the hydrostatic pressure at constant density $\rho_0$. We now introduce rescaled variables for space $\tilde{\mathbf{r}} = \mathbf{r}/H$, time $\tilde t= t D/H^2$, velocity $\tilde{\mathbf{u}} = \mathbf{u}H/D$, pressure $\tilde p = p H^2/ D \eta$ and nutrient concentration relative to its value $c_1$ in the absence of the colony $\tilde c = c/c_1$, where $H$ is the depth of the substrate fluid.
In the creeping flow regime, appropriate to our experiments, inertial terms on the left-hand side of the equation \eqref{boussi} can be neglected, and the governing equations then become (see Appendix \ref{sec:nondimensionalizing_flow} for details):
\begin{eqnarray}
\frac{\partial c}{\partial t} + \mathbf{u}\cdot \boldsymbol{\nabla} c &=& \nabla^2 c ,
\label{AD_solute}
\\
\nabla^2 \mathbf{u} - \boldsymbol{\nabla}p - \mathrm{Ra}\;c \mathbf{\hat{z}}&=& \mathbf{0},
\label{stokes_Ra}
\\
\boldsymbol{\nabla} \cdot \mathbf{u} &=& 0,
\label{incomp}
\end{eqnarray}
where the tildes have been dropped for convenience and $\mathbf{r}$, $t$, $\mathbf{u}$ , $p$ and $c$ now denote nondimensional variables. In Eq. \eqref{stokes_Ra}, the Rayleigh number $\mathrm{Ra} = h^3 \beta c_1 g / D \nu$, compares the buoyant forces to the stabilizing effect of the viscous forces.

We can consider again a colony with fixed radius $R$, provided the characteristic eddy turnover time $\tau_{\text{eddy}} \sim 1$ day for the baroclinic flow is much shorter than the characteristic radial colony growth time $\tau_{\text{growth}} = R(t) / \left( dR / dt \right) \sim 10 \ \text{days}$ for $t \gg t^\ast$. The colony expansion rate is slower than the induced flow velocity, and starts behaving like a solid in this regime, so we apply a no-slip boundary condition just below the colony. We also apply a no-slip boundary condition to the walls of the Petri dish and a free boundary condition to the air-substrate interface such that there is no normal velocity, $v_z=0$, and negligible shear stress, $\partial v_r/\partial z=0$. We apply the same nutrient absorption boundary condition below the yeast colony because the normal component of the fluid velocity at the boundary with the colony vanishes, and no-flux boundary conditions on both the walls of the petri dish and the fluid surface to the diffusing nutrient field.
We use OpenFOAM 5.0 \cite{openfoam} to solve the governing Equations \eqref{AD_solute}-\eqref{incomp} with the boundary conditions given by Eq. \eqref{yeast_bc}, using the program \texttt{stokesBuoyantSoluteFoam} \cite{stokesBuoyantSoluteFoam} (available on GitHub) with the experimentally measured parameters in Table \ref{tab:model_parameters}; see Appendix \ref{sec:simulation_methods} for additional numerical details. 

The baroclinic effect leads to an intense vortex ring beneath the outer edge of the colony, as revealed by the transverse section shown in Fig. \ref{simu_flow} a).
The flow geometry and intensity on the surface of the fluid resemble the experimental flow field shown in Fig. \ref{piv} around the colony. As shown in Figure \ref{simu_flow}b), the corresponding radial velocity profile at the fluid interface is in good agreement with the experimental profile, with a strong peak at about 1.5 times the colony radius. Figure \ref{simu_flow} c) compares the maximum radial velocity measured in the stationary flow regime, reached after 48 hours in the experiments, with simulations as a function of the substrate fluid height $H$. Our minimal buoyant flow model tracks the experimental peak velocities, supporting the hypothesis of a buoyancy-driven flow produced by a baroclinic instability in our experiments.

\section{Model Coupling Growth with Dilational Flow  \label{sec:model_of_dilation}}

\begin{figure*}
\includegraphics[width=17.8cm]{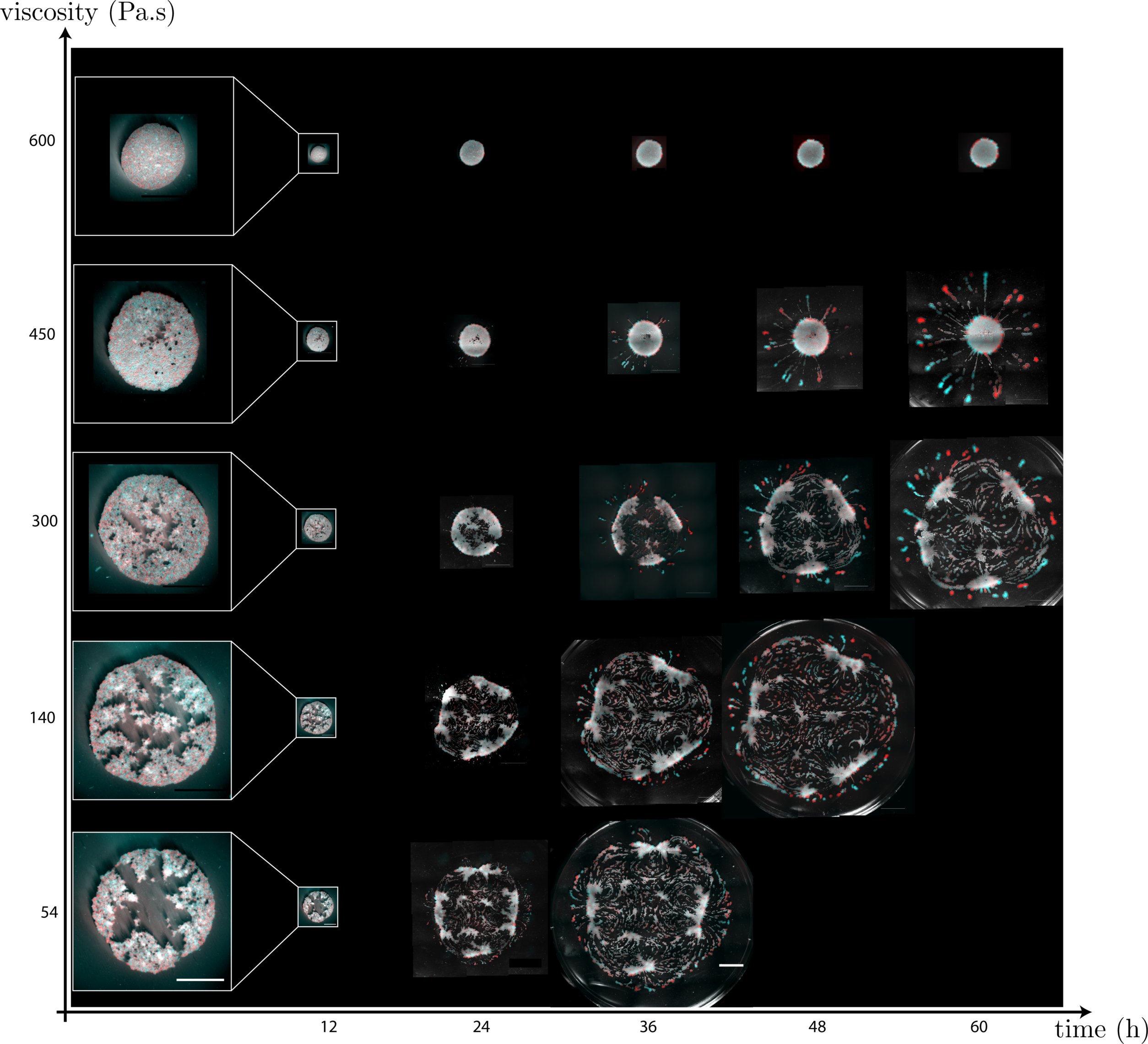}
\caption{Morphologies of yeast colonies growing on a liquid media substrate over time at a variety of viscosities. Quoted substrate viscosities are accurate to about 10\% (see Table \ref{tab:viscosity_table}). The figure shows merged brightfield and fluorescent images. White: transmitted brightfield, red: YFP strains, and cyan: mCherry strains. All images have the same scale and the scale bar at the lower right corresponds to 10 mm. The left column is an enlargement on the colonies after 12h of growth and its scale bar corresponds to 5 mm.
\label{PD}}
\end{figure*}

In this Section, we investigate how substrate viscosity influences colony morphology and describe a simple phenomenological model for colony growth, expansion, and thinning in the spirit of the so-called lubrication approximation \cite{batchelor1967_book}. Figure \ref{PD} displays five characteristic colony morphologies over time growing on liquid media for the entire range of studied viscosities (Table \ref{tab:viscosity_table}), from $\eta = 54 \pm 8$ Pa$\cdot$s  to $\eta = 600 \pm 90$ Pa$\cdot$s. Our measurements of flow velocity shown in Figure \ref{piv} reveal that metabolically driven buoyant flows become apparent as early as 2 hours after inoculation, suggesting that yeast cells can deplete enough mass to induce a flow even at this initial stage of growth. The first column to the left on Figure \ref{PD}  shows an enlargement of the colonies 12h after inoculation. Their shape already shows a strong dependence to the substrate viscosity, suggesting that the future morphology of the colony is determined during the early growth.

When viscosity decreases, the amplitude of the toroidal flow field beneath the colony increases and eventually applies enough force to alter the initial circular morphology of the colony. For instance, experiments performed at $\eta = 54 \pm 8$ Pa$\cdot$s indicate that the flow velocity can reach magnitudes up to $20$ mm/day and apply non-negligible stresses on the colony.
Once the cell division rate falls behind the colony's advancing front at $t \approx t^\ast$, the bulk of the colony ceases to behave as a liquid with internal motion due to cell division and begins to behave more like a viscoelastic material. Given the much faster fluid substrate velocities outside the colony relative to the colony expansion speed, the colony starts experiencing radial shear stresses imposed by the flow. 
One possible explanation for the especially intriguing colony morphology displaying multiple elongated fingers around the colony edge, close to $\eta = 450 \pm 70$ Pa$\cdot$s, could be a mechanism similar to viscous fingering instabilities. Under these conditions, competition between relaxation forces, due to the attractive interaction between cells and an outward pulling force produced by the radial flow could drive an instability resembling those that arise in rotating oil films \cite{melo1989a}, suitably modified to allow for colony growth and the discreteness of the underlying cells.
However, when the viscosity drops below $\eta \lesssim 300 \pm 45$ Pa$\cdot$s, the radial expansion imposed by the vortex ring under the colony starts to outcompete the colony expansion due to cell divisions, such that growth cannot accommodate the dilational flow during the initial stage. This results in a rapid separation of the cells and holes start opening up within the center of the colony.

A complete understanding of the complex experimental behaviors described here (exponential stretching prior to genetic demixing, a fingering instability with fingers that break into droplet-like clusters and fragmentation; see Fig. 1) would require a detailed theory of the fluid dynamics of the substrate fluid coupled to the visco-elastic behavior of a colony of approximately 5-micron-sized cells with both excluded volume and attractive interactions, all while cells are actively dividing, as well as interacting with the substrate fluid during the range expansion. We hope that the results described here will encourage such theoretical investigations, which might also need to account for the discreteness of the cells in the colony and assess the impact of the fluid mechanics on the genetic demixing observed in our experiments.

Here, we propose, instead, a simple phenomenological model that provides insight into the exponential stretching and colony thinning during the early stages of the range expansion when the colony maintains its circular symmetry and behaves approximately like a two-dimensional liquid. In analogy with treatments of colony expansions on hard agar plates \cite{wakita1994}, we describe the dynamics of the colony height by a generalized Fisher population dynamics equation \cite{murray_spatial_book_2011} for the colony height $h(\mathbf{r},t)$, namely:
\begin{equation}
\begin{aligned}
\frac{\partial h(\mathbf{r},t)}{\partial t}     &+ \boldsymbol{\nabla} \cdot \left[ h(\mathbf{r},t) \mathbf{v}(\mathbf{r}) \right]\\
                                                                        &= D_h \nabla^2 h(\mathbf{r},t) + \mu h(\mathbf{r},t) \left[ 1 - \frac{h(\mathbf{r},t)}{h_0} \right],
\label{model}
\end{aligned}
\end{equation}
where $\mathbf{v}(\mathbf{r})$ is the advecting hydrodynamic flow velocity that acts on the colony and $\mu$ is an effective colony vertical growth rate when its height is small. The quantity $h_0$ is the steady state colony thickness in the absence of flow and spatial gradients of the height field, which we expect will depend on quantities such as nutrient penetration depth inside the colony \cite{Lavrentovich2013} and strength of, e.g. , the Van der Waals and gravitational forces that attract the cells to the liquid substrate. The parameter $D_h$ is a diffusion constant that promotes an approximately uniform colony height -- a similar term appears in, e.g., the hydrodynamic equations that describe capillary wave-like excitations in thin helium films \cite{bergman1971}. 


One source of the radially outward flows we observe near the surface during the early stages of our range expansions on liquid substrates is the outward pushing by the growing quasi-two-dimensional yeast colony. To determine the form of this contribution to the substrate flow, we assume that, at least during the early stages of the expansion, the colony behaves like a two-dimensional liquid where all the cells in the colony receive enough nutrients to actively divide. We further assume that the two-dimensional colony viscosity can be neglected compared to the overdamped frictional coupling to the liquid substrate. We can then apply a simple hydrodynamic model, \cite{klapper2002,head2013,farrell2013,giometto2018}, which leads to:
\begin{equation} \label{div_growth}
\nabla^2 p_{2d} = -\gamma \boldsymbol{\nabla}\cdot \mathbf{v} = - \gamma \alpha_{1}
\end{equation}
where $p_{2d}$ is an effective two-dimensional pressure field inside the colony \cite{giometto2018}. Here, $\alpha_{1}$ arises from cell divisions that, as shown below, will give rise to a horizontal radial velocity field within the quasi-two-dimensional liquid colony averaged over the thickness of the colony. The quantity $\gamma$ is a frictional coefficient due to the motion of the colony relative to the liquid substrate. If the liquid substrate has a dynamical viscosity $\eta_s$ and depth $H$, in the limit of colony radius larger than $H$, we then expect $\gamma \approx \eta_s / hH$ \cite{stone1998}, where $h$ is the thickness of the colony. We can now exploit an electrostatic analogy, such that the two-dimensional pressure field inside the colony satisfies a Poisson equation, and where the height-averaged growth rate $\alpha_{1}$ determines a 2d ``charge density''.  The colony velocity field (like the 2d electric field inside a charged disk in two dimensions) that solves Eq. \eqref{div_growth} has the radially symmetric form:
\begin{equation} \label{Efield}
\mathbf{v}(x,y) = \frac{1}{2} \alpha_{1} \; r \mathbf{\hat{r}}, \; r = \sqrt{x^2+y^2}.
\end{equation}
When coupled to an underlying viscous substrate fluid, this dilational flow field within the colony will act to induce flows in the underlying liquid, in qualitative agreement with our PIV measurements near the surface shown in Fig. \ref{piv}(a). 
\begin{figure}
\includegraphics[width=8.6cm]{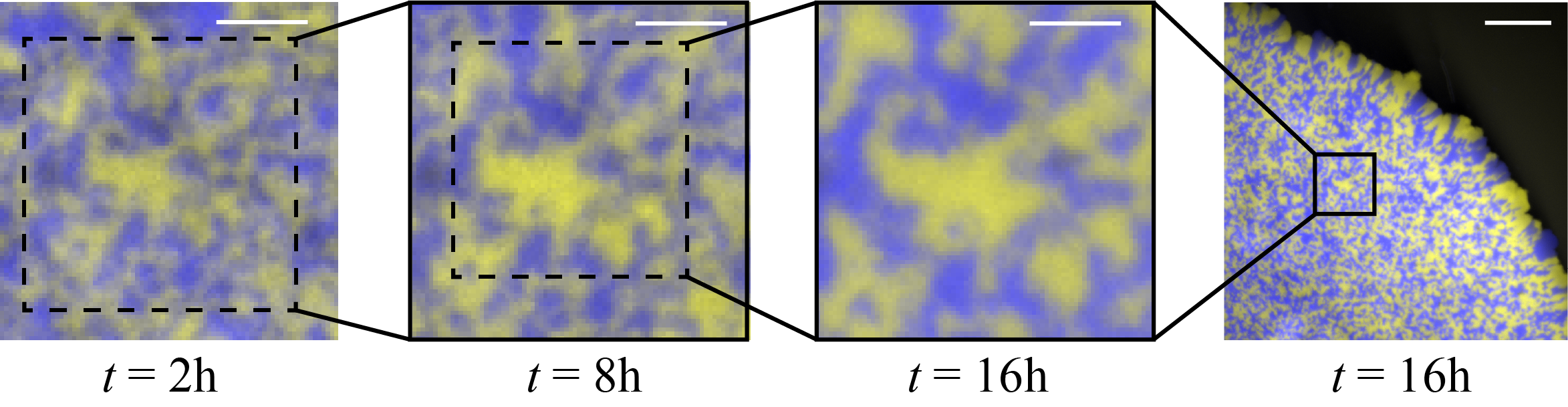}
\caption{Magnification of  demixing patterns formed by two different yeast strains growing on a substrate with a viscosity $\eta = 600 \pm 90$ Pa$\cdot$s at different time points. The first three images have the same scale represented by the white bar on the upper right of the images; the scale bar corresponds to 100 $\mu$m. The final picture to the right shows the same feature at the larger colony scale; the scale bar now corresponds to 500 $\mu$m.
\label{hodag}}
\end{figure}
The development of expanding genetic patterns during the approximately exponential growth for $t<t^\ast$ is shown in Fig. \ref{hodag}. The figure highlights one particular feature inside the black dashed square which only undergoes a dilatation when expanding over time, as if the genetic patterns were painted on the surface of an inflating balloon, which is also consistent with Eq. \eqref{Efield}. Estimates of this dilational expansion velocity for $t<t^\ast$ gives values of the order of $4-8$ mm/day, the same order of magnitude as the colony front expansion velocity observed during the early exponential expansion regime.

\begin{figure}
\includegraphics[width=7cm]{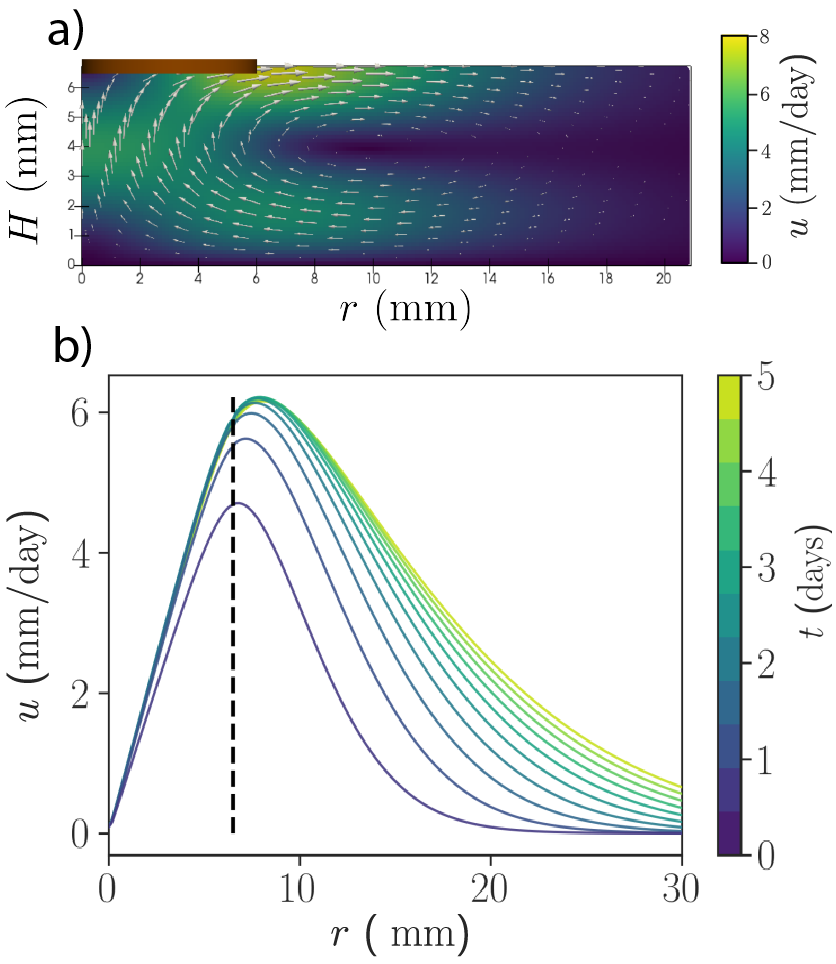}
\caption{a) Snapshot of the simulated flow field below the yeast colony (brown bar) after flow is initiated when $t \gg t^\ast$. The simulated flow field is very similar to the one displayed in Fig. \ref{simu_flow}a) except with free boundary condition beneath the yeast colony. b) Azimuthal average of the numerical flow field using the measured parameters in Table \ref{tab:model_parameters} plotted every 12 hours at the substrate fluid surface with free boundary condition beneath the yeast colony. 
\label{freeBC}}
\end{figure}

The second source of the flow we need to account for is the more vigorous motion driven by the baroclinic instability. This flow is present for $t > t^\ast$, and becomes dominant at later growth stages for high substrate viscosity, and at increasingly earlier times with decreasing substrate viscosity.
Triggered by the metabolic uptake of nutrients, this additional flow is potentially responsible for the fingering and fragmentation instabilities observed when the substrate viscosity decreases and flow amplitude becomes larger.
If we express the flow produced by cell divisions occurring throughout a circular colony in the form $\mathbf{v}_1(\mathbf{r}) = \frac{1}{2} \alpha_1 r \hat{\mathbf{r}}$, we expect then another contribution to this velocity of the form $\mathbf{v}_2(\mathbf{r}) = \frac{1}{2} \alpha_2 r \hat{\mathbf{r}}$ once the baroclinic instability establishes a vortex ring in the substrate fluid beneath the colony with a size of order 1.5 times the colony radius (Figure \ref{piv}). A simple model of a vortex ring submerged in substrate fluid with an image vortex ring with opposite circulation above the colony satisfies the requisite boundary conditions beneath the colony (the resulting velocity field resembles the magnetic field from a pair of anti-Helmholtz coils). This ansatz leads to a radial velocity field at the colony which vanishes linearly in $r$ for small $r$, and falls off roughly like $1/r^4$ for $r$ large compared to the colony radius.
To check these ideas for the substrate-induced velocity field acting on the colony, we have repeated the simulations of Sec.\ref{sec:hydrodynamic_simulations}.B under identical conditions with, however, free instead of no-slip boundary conditions at the interface between the colony and the substrate fluid.  We thus assume that active cell divisions throughout a circular colony cause it to behave like a two-dimensional liquid, with a contribution to the in-plane colony velocity field imposed directly by the substrate fluid. The resulting flow snapshot for the substrate fluid velocity field below the colony, displayed in Fig. \ref{freeBC}a), is qualitatively similar to Fig. \ref{simu_flow}a) indicating a submerged vortex ring. Now, however, the absence of a no-slip boundary condition leads to a velocity field right at the colony-substrate interface. The azimuthal average of our numerical flow field is shown in Fig. \ref{freeBC}b), again at 12 hour time intervals. The results are similar to Fig. \ref{simu_flow}b), except that they clearly show a linear behavior of the velocity field underneath the colony, consistent with the ideas in the preceding paragraph.

With these motivations, it seems reasonable to assume that the advecting velocity field in Eq. \eqref{model} takes the form:
\begin{equation}
\mathbf{v}(\mathbf{r}) = \frac{1}{2} \alpha r \hat{\mathbf{r}}
\label{eq:radial_velocity_and_alpha}
\end{equation}
where $\alpha$ is an effective dilational flow parameter that includes the effect of the baroclinic instability as well pushing generated by dividing cells within the colony. We expect $\alpha$ to increase with decreasing substrate viscosity, reflecting a stronger baroclinic instability.

With these assumptions, Eq. \eqref{model} takes the form:
\begin{equation}
\begin{aligned}
\frac{\partial h(\mathbf{r},t)}{\partial t}     &+ \frac{1}{2} \alpha r \mathbf{\hat r} \cdot \boldsymbol{\nabla} h(\mathbf{r},t)\\
                                                                        &= D_h \nabla^2 h(\mathbf{r},t) + (\mu-\alpha) h(\mathbf{r},t)  - \frac{\mu h^2(\mathbf{r},t)}{h_0}.
\label{model_2}
\end{aligned}
\end{equation}
In regions where the colony height is spatially uniform, we have for the height $h(t)$, $\frac{\partial h(\mathbf{r},t)}{\partial t} = (\mu-\alpha) h(\mathbf{r},t)  - \mu h^2(\mathbf{r},t) / h_0$, and thus:
\begin{equation}
h(t) = \frac{h(0) e^{(\mu - \alpha)t}}{1+\frac{\mu h(0)/h_0}{\mu - \alpha} \left( e^{(\mu - \alpha)t} -1 \right)} .
\label{h_sol}
\end{equation}
We can now look for a radially symmetric solution with an interpolating step-like function $\Theta(x)=1, \; x \ll 0, \; \Theta(x)=0, \; x \gg 0$,
\begin{equation}
h(r,t) = h(t) \Theta \left[( R(t) -r \right) / \delta],
\end{equation}
where $R(t)$ defines a colony radius smeared out over an interfacial width $\delta$. It is easy to see from Eq.\eqref{model_2} that, provided $r \gg \delta$ and $r \gg \sqrt{D_h/\alpha}$, the colony radius grows exponentially in time:
\begin{equation}
R(t)=R(0)e^{\frac{1}{2}\alpha t}.
\label{R_sol}
\end{equation}

\begin{figure}
\includegraphics[width=8.6cm]{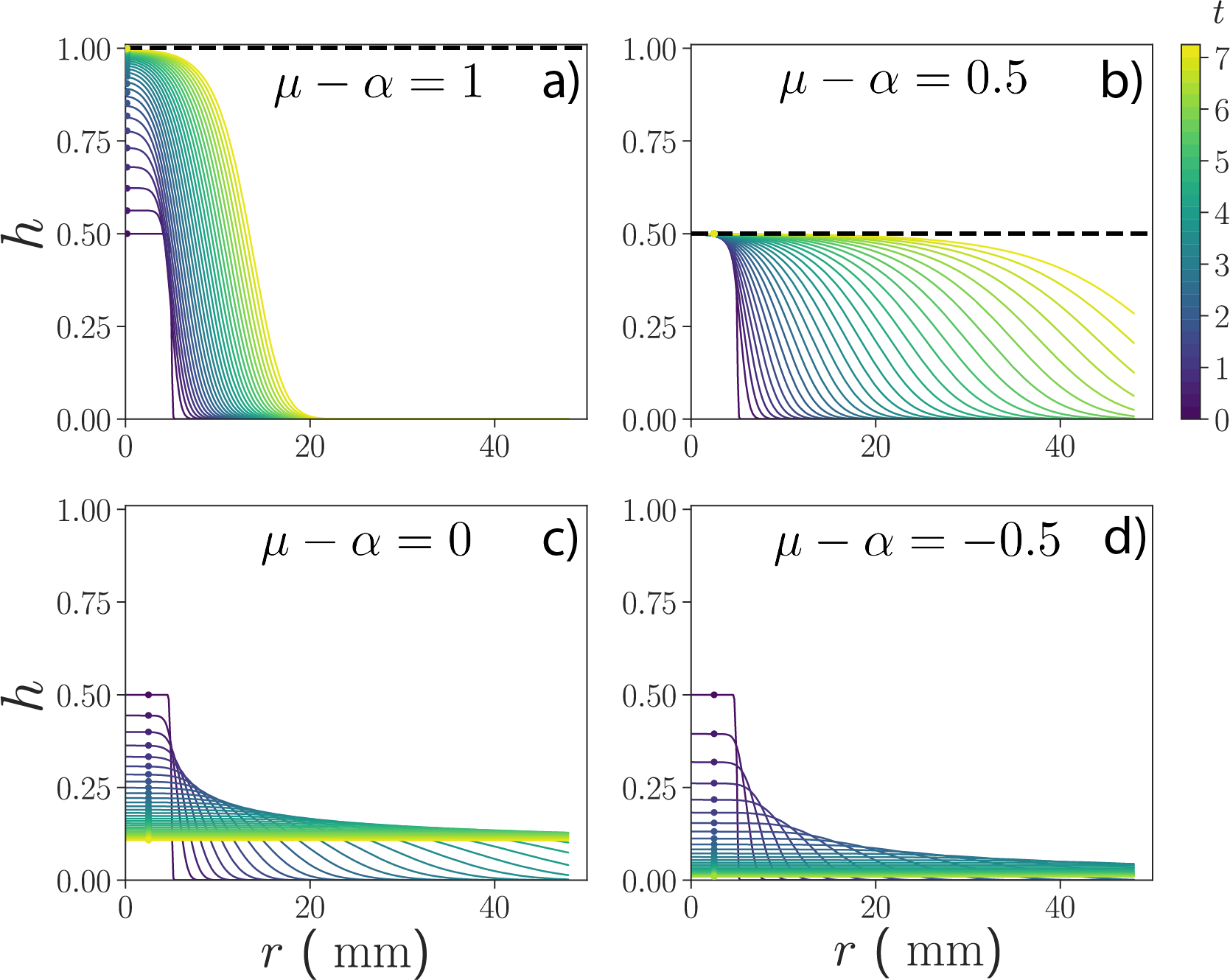}
\caption{Numerical solution of Eq. \eqref{model} for $h(\mathbf{r},t) / h_0$ at different values of $\mu - \alpha$ and equal time intervals. The radial coordinate $r$ is measured in units of $\sqrt{D_h / \mu}$, the width of the Fisher wave  in the absence of a dilational flow. The colored dots correspond to the prediction
of $h$ as a function of time from equation \eqref{h_sol}; it is clear that
there is good agreement between the theoretical prediction and simulation. a) for $\alpha=0$ the colony height increases to $h/h_0=1$ and the front propagates radially with a constant velocity $v_\text{F} = 2\sqrt{D_h \mu}$ with $\mu=1$. b) When $\alpha <\mu$ the colony front propagation velocity increases exponentially with time and the colony height decreases to $h^\ast = h_0 (1 - \alpha / \mu) < h_0$. c) When $\alpha=\mu$ the dilational flow is strong enough to decrease the colony height below one cell size and $h(t)$ goes to zero logarithmically with radius. d) For $\alpha>\mu$, the colony thins exponentially fast, potentially
signalling that holes open during its early exponential growth; these holes may be responsible for the highly fragmented colonies at later times.
\label{h_simu}}
\end{figure}

Figure \ref{h_simu} shows the numerical solution of equation \eqref{model}, assuming radial symmetry for the colony height $h(\mathbf{r},t) = h(r,t)$, at different values of $\alpha/\mu$ using the program \texttt{forcedThinFilmFoam} \cite{forcedThinFilmFoam}; see Appendix \ref{sec:simulation_methods} for more details. In the absence of an advecting velocity field, $\alpha=0$ in Fig. \ref{h_simu}a), Eq. \eqref{model_2} has the usual Fisher wave solution of an outwardly expanding colony front circumference with constant velocity $v_\text{F} = 2\sqrt{D_h \mu}$ whenever the colony radius is much greater than the interfacial width $l_\text{F} = \sqrt{D_h / \mu}$ \cite{murray_spatial_book_2011}.
However, for nonzero $\alpha$ such that $\mu-\alpha>0$ in Fig. \ref{h_simu}b), we find an \textit{exponentially} fast advance of the wave: if the shoulder of the population wave in this case occurs at $x_0$ when $t=0$, then the position of the shoulder at time $t$ is as $x_0 \exp\left[(1/2)\alpha t\right]$, with a width $\delta$ of order $\sqrt{D_h/(\mu - \alpha)}$, consistent with our early time observations in Fig. \ref{fronts}. In this regime, the colony advances but is thinned down to a height given by the long time limit of Eq\eqref{h_sol}:
\begin{equation}
h^\ast = h_0 \left( 1 - \frac{\alpha}{\mu}\right).
\end{equation}
Thus, with increasing $\alpha$, the flow becomes stronger and the exponential advance of the colony is faster but the colony becomes progressively thinner.
Interestingly, when $\alpha=\mu$, equation \eqref{h_sol} becomes $h(t) = h(0) / \left( 1+ \frac{h(0)}{h_0}\mu t \right)$, and approaches zero as $h(t)\approx h_0/(\mu t)$ for large times. In fact, when time is substituted with $R$ using
equation \eqref{R_sol}, we find that at large times the height at the midpoint of the shoulder behaves according to
\begin{equation}
h_s[R(t)] \sim  \frac{h_0}{\ln\left[\frac{R(t)}{R(0)}\right]}
\end{equation}
such that $h$ decreases logarithmically with radius, leading to the formation of a wide plateau due to the extremelly slow decay of $h$ over time, as can be seen in Figure \ref{h_simu}c).  
For sufficiently strong flows such that $\alpha>\mu$, there is a ``thinning catastrophe'', see Figure \ref{h_simu}d), such that the colony population collapses at long times. In this limit, of course, the discrete nature of the cells making up the colony, neglected in Eqs \eqref{model} and \eqref{model_2}, becomes important.
 
\begin{figure}
\includegraphics[width=8.6cm]{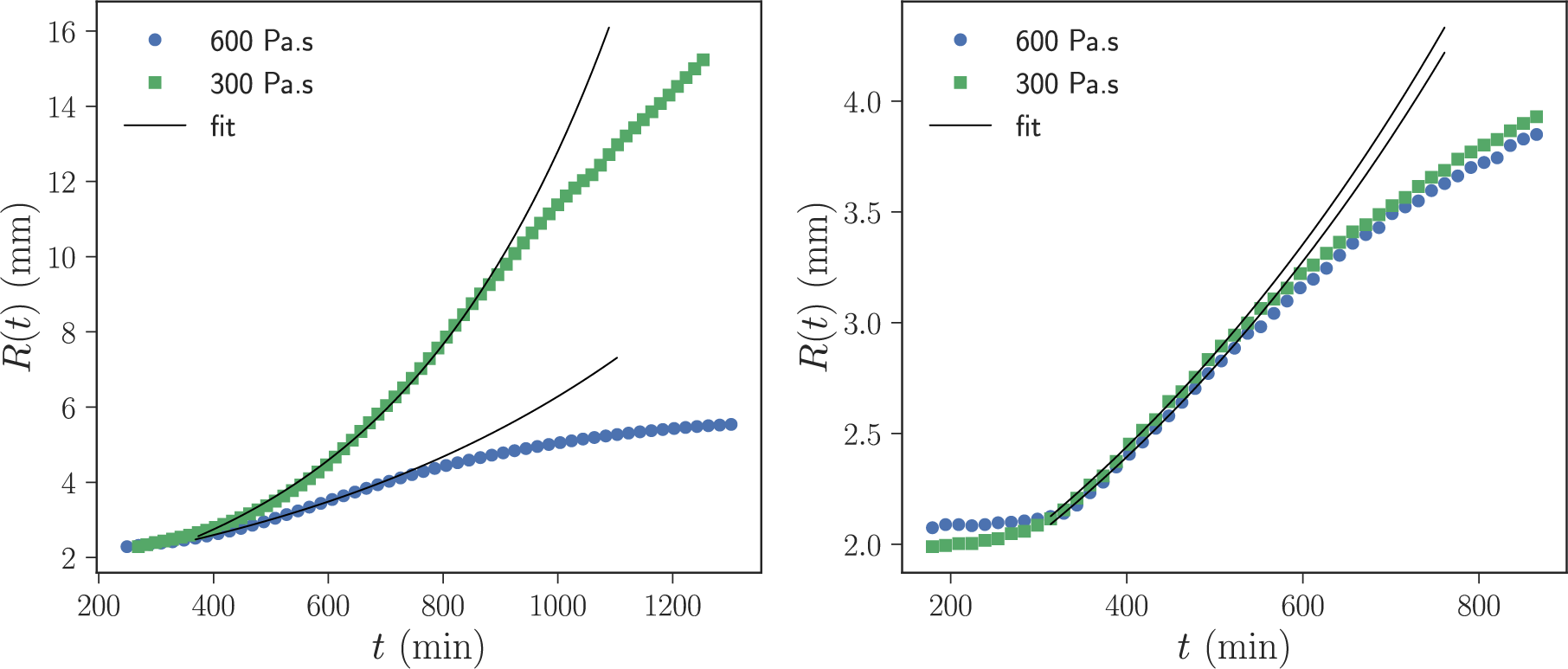}
\caption{a) Colony radius as function of time during the first day of growth for two different viscosities; blue circles, $\eta = 600 \pm 90$ Pa$\cdot$s, green squares: $\eta = 300 \pm 45$ Pa$\cdot$s and black line: exponential fit realized for $t<t^\ast$. The short time behavior is consistent with an exponential growth of the colony radius in both cases, but the growth is much faster at lower viscosity. b) same as in a) with experiments realized for colonies growing on a 1 mm thin substrate liquid film on the top of a nutrient rich gel layer. We found that the exponential fit realized for $t<t^\ast$ exhibit a similar expansion rate for both viscosities.
\label{expansion}}
\end{figure}

Finally, we check the qualitative agreement between this simplified model and the experiments by determining the colony expansion rate during the superlinear growth regime ($t<t^\ast$) as a function of substrate viscosity. A detailed measurement of the radial expansion coefficient's viscosity-dependence, $\alpha = \alpha(\eta)$, would provide a more quantitative test. Here, we explore this idea further by reproducing the same experiments as the ones described in Sec. II for two different substrate viscosities: by relating the above model predictions to the early colony morphologies, one may be able to estimate a critical viscosity below which the flow becomes strong enough to cause a ``thinning catastrophe''. As can be seen in the first column of Fig. \ref{PD}, for $\eta \lesssim 300 \pm 45$ Pa$\cdot$s, holes start opening up in the center of the colonies during the early expansion, indicating that the flow dilation rate is larger than the colony height growth rate and corresponds to the height profile regime described by $\alpha>\mu$. Assuming $\eta \simeq 300 \pm 45$ Pa$\cdot$s is the highest viscosity at which we can observe a catastrophic thinning of the colony height in the early growth regime, our model suggests that $\alpha \approx \mu$ in this experiment. 

Figure \ref{expansion}a) displays the colony radius $R(t)$ over time growing on two different liquid substrates with viscosity $\eta = 600 \pm 90$ Pa$\cdot$s and $\eta = 300 \pm 45$ Pa$\cdot$s. The colony expansion rate described by Eq. \eqref{R_sol} can then be estimated from an exponential fit of $R(t)$ for $t<t^\ast$, and gives $\alpha = 4.2 \pm 0.4 \; \mathrm{day}^{-1}$ for the higher viscosity, and a larger rate $\alpha = 7.2 \pm 0.4 \; \mathrm{day}^{-1}$ for the lower viscosity. Assuming that the critical value of $\alpha$, for which we have $\alpha \approx \mu$, is close to the colony expansion rate measured for $\eta = 300 \pm 45$ Pa$\cdot$s, we estimate $\mu \approx 7.2 \pm 0.5  \; \mathrm{day}^{-1}$, which  gives a characteristic division time of $\tau \approx 140$ min in the vertical direction of the colony, in approximate agreement with yeast colony growth rates on hard agar plates \cite{giometto2018}.

The dilational coefficient $\alpha$ in Eq. \eqref{eq:radial_velocity_and_alpha} is presumably a combination of the $\alpha_1$ and $\alpha_2$ contributions discussed above. Although it is difficult to determine the value of $\alpha_2$ for $t<t^\ast$, as the metabolic velocity field is weaker at short times, we were able to isolate the constant $\alpha_1$, related to the flow contribution coming from cell-divisions at a liquid interface but \textit{without} the enhanced dilational velocity due to the metabolic flow. To do this, the same experiments were repeated on a much thinner 1 mm thick layer of liquid substrate deposited on the top of a regular, nutrient rich gel plate. This geometry allowed us to damp out the baroclinic instability in the thin liquid layer, and revealed a nearly identical expansion rate this time, with $\alpha = 4.2 \pm 0.3\; \mathrm{day}^{-1}$ for both $\eta = 300 \pm 45$ Pa$\cdot$s and $\eta = 600 \pm 90$ Pa$\cdot$s, suggesting that $\alpha_1$ is independent of substrate viscosity for $300 \leq \eta \leq 600$ Pa$\cdot$s.
Note that the measured value of $\alpha_1$ is similar to the expansion rate $\alpha$ we found for thicker substrates at higher viscosity, while it is significantly less than the measured $\alpha$ for the substrate with lower viscosity. This suggest that the metabolic flow doesn't contribute significantly to the colony expansion for $\eta = 600 \pm 90$ Pa$\cdot$s, while it considerably increases the colony dilation rate for $\eta = 300 \pm 45$ Pa$\cdot$s even at early times for $t<t^\ast$.
Although further experiments would be required to fully map out the colony dynamics as a function of substrate thickness and viscosity, our experimental results suggest a qualitative agreement with equations \eqref{model} and \eqref{model_2}.

\section{Discussion \label{discussion}}

We investigated the growth of yeast range expansions on the surface of an extremely viscous nutrient-rich liquid substrate. Capillary forces keep our yeast cells at the surface for many days, and the extreme viscosity of the fluid insures that cell clumps that break the surface of the air-liquid interface settle slowly. The large viscosity also prevented thermal convection from mixing the media. Previous experiments of range expansions on solid agar media featured a thin layer of proliferating cells at the frontier of radially expanding circular colonies \cite{Hallatschek2007}. We found that colonies grown on a liquid medium, where the substrate can flow and friction between the cells and the medium is much lower, behave very differently. 

In the early stages of these range expansions, for $t<t^\ast$, colony radii grew in a superlinear, approximately exponential fashion and the growth was dominated by active cell divisions throughout the colony. However, for $t>t^\ast$, yeast metabolism generated fluid flows in the surrounding media many times larger than their basal expansion velocity. This flow dramatically altered the colony morphology, depending on the surrounding substrate viscosity.

Compact circular colonies grew for $\eta \approx 600\pm90$ Pa$\cdot$s (3.0\% polymer), the largest viscosity we tested, featuring a regime of roughly exponential stretching and thinning where strains remained mixed together, and later a period of slow, linear expansion where strains genetically demixed and resembled expansions on agar plates \cite{Hallatschek2007} with more wiggly domain walls. The expansion likely slowed because of nutrient depletion.

As the viscosity of the medium decreased, hydrodynamic forces acting on the colony were eventually sufficient to produce fingering and fragmentation instabilities and led to two additional morphologies.  At intermediate viscosities between  $\eta = 450\pm70$ Pa$\cdot$s and $\eta=300 \pm 45$ Pa$\cdot$s  (2.8\% -- 2.6\% polymer), compact colonies developed ``fingers'', an instability that allowed thin streams of cells to be ripped away from colonies resembling dendritic crystal growth in the presence of a solute-driven buoyant flow \cite{Steinbach2009} or fingering instabilities in spinning drops \cite{melo1989a} and Marangoni flow \cite{Keiser2017}. We attribute this liquid-like behavior at the colony perimeter to the lubricating effect of active cell divisions. The filaments then broke into clusters via a process reminiscent of capillary forces in the Raleigh-Plateau instability \cite{Schwartz1997,Waitukaitis2011}, with, however, differences due to actively dividing, discrete cells. The competition between the self-induced flow, diffusion of nutrients and the attractive forces between the cells might trigger a selection for a characteristic finger width.

\begin{figure}
\includegraphics[width=8.6cm]{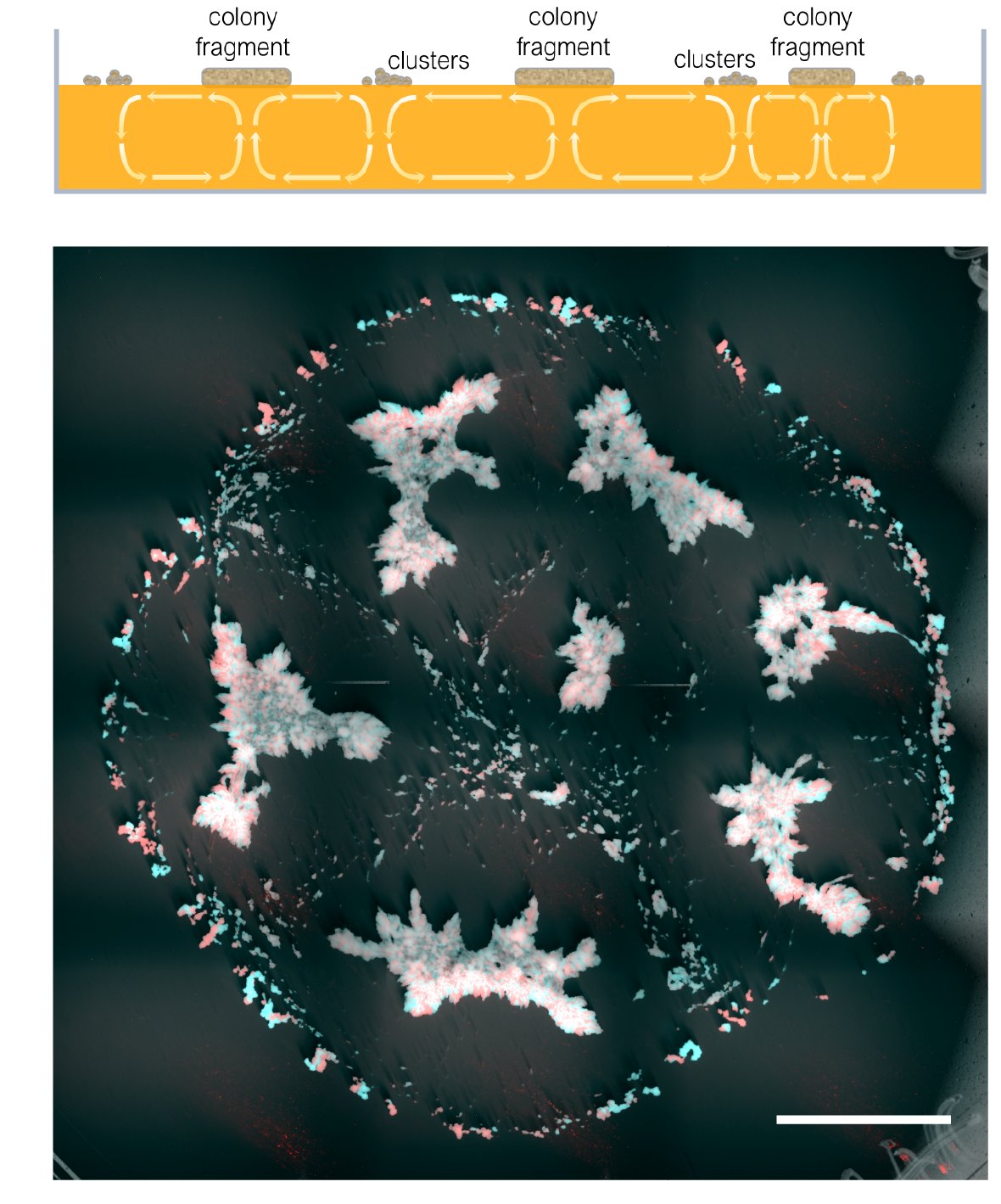}
\caption{Low viscosity ($\eta = 300\pm45$ Pa$\cdot$s) range expansion on a liquid substrate in the fragmentation regime.  This image was taken for $t \gg t^*$ in a single experiment under conditions similar to those in Fig.\ref{examples}d), except that the substrate fluid height was $H= 4$mm instead of 7mm. The more isolated cell fragments clearly collect on the mid-planes separating the larger ``continents",  consistent with the down-wellings associated with a vortex ring underneath each continent, as suggested by the sketch on the top. The scale bar corresponds to 10 mm.
\label{pangaea}}
\end{figure}

For viscosities lower than $\eta = 300\pm45$ Pa$\cdot$s, growing colonies exhibited solid-like behavior in the interior; they fractured into many irregularly shaped repelling island-like fragments. These repelling fragments could colonize an entire Petri dish within 36 hours, presumably because each fragment metabolically generated its own submerged vortex ring. This conjecture about a vortex ring under each solid-like colony fragment is consistent with the image shown in Fig. \ref{pangaea}, taken under experimental conditions similar to Fig. \ref{examples}d), but with a shallower substrate fluid. As opposed to the nearly monoclonal fingers separating from the initial colony after demixing, island-like fragments tended to be genetically diverse as the entire colony broke apart.

Our experiments and simulations provide strong evidence that yeast metabolism creates fluid flow in the surrounding media via a baroclinic instability: yeast created a pocket of less dense fluid \textit{on top} of a more dense one that generated vorticity near the colony edge when the isobars and isoclines of the underlying fluid crossed each other at an angle. Minimal buoyant fluid flow simulations calibrated to experiments with independently measured parameters capture our experimentally observed flow fields. Interestingly, as discussed in Appendix\ref{sec:calibrating_simulation}, these calibrations allowed us to measure the mass flux rate into the yeast colony in rich nutrient conditions as $a c_0 = 5 \pm 2 \ \mathrm{pg/(\mu m^2 \ hour)}$; the authors are unaware of other literature measuring this quantity. Furthermore,
this mass flux rate is consistent with a nutrient screening length of about
50 $\mathrm{\mu m}$ inside yeast colonies (Appendix \ref{sec:calibrating_simulation}), consistent with that measured in prior work \cite{Lavrentovich2013}. 
 
Furthermore, colonies always generated fluid flows against the direction of gravity, regardless of their position in a sealed chamber, and we found that yeast cells grown to saturation in overnight culture decrease the surrounding media's density by $\Delta \rho = -0.0090 \pm 0.0005  \ \mathrm{g/mL}$. We believe that surface tension gradients (the Marangoni effect) played only a minor role in generating the observed flows, because yeast attached to the surface of a sealed chamber generated fluid flow comparable in magnitude, and because the above arguments suggest that buoyancy alone sufficiently explains the phenomenon. To the best of our knowledge, this unusual baroclinic instability has not been previously investigated in a biological context.

The work described here suggests a number of intriguing avenues for future work:
for example, can other microorganisms growing on or near the surface of liquids generate buoyant flows similarly to our experiments? Preliminary experiments with immotile \textit{E.\@ coli} colonies  have exhibited similar flows when growing on the surface of liquid substrates with comparable viscosity, and have also exhibited fascinating colony morphologies \cite{[{See also }]weinsteinThesis_2018}.
It is intriguing to speculate that similar instabilities might occur at much higher Reynolds numbers in the oceans, beneath plankton blooms confined to, say, the first 50 meters of depth.
It would also be interesting to experimentally test if microbial colonies that generate buoyant flows have a selective advantage relative to those that do not. Induced fluid flows clearly allow more efficient redistribution of nutrients and provide a mechanism for the more rapid dispersal of colony fragments. Preliminary numerical investigations when viscosity is lowered from infinity (i.e.\ modeling hard agar substrates), increasing the Rayleigh number from 0 to $10^4$ in our Petri dish geometry, increased the nutrient absorption rate of the yeast colony by a factor of about 1.5, suggesting that colonies generating stronger buoyant flows could indeed have a selective advantage (see Appendix \ref{sec:nutrient_absorption_buoyant_flow} for details).

Although yeast colonies might develop fluid-mechanics-like instabilities reminiscent of classical ones in the presence of flow \cite{melo1989a, Schwartz1997, Eggers1997a, Keiser2017}, they differ in two key ways: 1) Dividing cells cause \textit{growth} over time, stressing the need for further theoretical work to understand instabilities arising from the competition between flow and growth; and 2) the discreteness of the dividing cells may play an important role near the ``thinning catastrophe'' discussed for a simple theoretical model in section \ref{sec:model_of_dilation}. The transition from an approximately exponential to a slower expansion rate, corresponding to the transition from liquid-like to solid-like behavior of the yeast colony, could also benefit from a fluid-mechanical perspective to model the yeast fingering instability, assuming a liquid-like behavior due to agitation by cell-divisions at the frontier.

The origin of the \textit{quantitative} differences between yeast colony growth on the highest viscosity substrates and on hard agar plates, such as the more wiggly genetic domain boundaries has yet to be understood. Systematic investigations of how colony morphology and genetic patterns vary with nutrient concentration (glucose) in addition to viscosity, similar to the pioneering work of Wakita et al.\@ \cite{wakita1994}, would also be of interest.
Furthermore, it is worth noting that we modeled the rheology of the liquid substrate as a Newtonian fluid despite the shear-thinning properties measured in the media at very large polymer concentrations as discussed in Appendix \ref{rheo}; future work should investigate how more pronounced non-Newtonian effects could impact the fluid flows induced by the yeast, in the context of microbial populations growing in mucus for instance \cite{li2013}.

Lastly, the fluid used in this paper is viscous enough that it can be advected at a velocity as low as $1 \ \mathrm{mm/day}$ \cite{[{See also }]weinsteinThesis_2018}, matching the expansion rates of \textit{E.  coli} and the baker's yeast \textit{S. cerevisiae} on agar  \cite{Weinstein2017,Korolev2012a}, over an entire  $10 \ \mathrm{cm}$ Petri dish over many days of growth \cite{[{See also }]weinsteinThesis_2018}.  The extreme viscosity of the fluid allows for the imposition of slow, controlled fluid flows at a macroscopic scale that can advect microbial colonies and provides an alternative to working with microfluidic devices where  complications arise when microbes  stick to the walls of their enclosure \cite{[{See also }]weinsteinThesis_2018,Tesser2017}.
Using syringe pumps, one could impose well-defined flows on microbial colonies and systematically repeat previous experiments with microbial range expansions on hard agar plates \cite{Korolev2012a,Gralka2016b,Weinstein2017,Muller2014,Weber2014,McNally2017,Lavrentovich2016} on viscous liquid substrates like those studied here but with additional types of advection. Investigating the evolutionary dynamics of colonies composed of complementary strains that secrete public goods such as leucine and tryptophan \cite{Muller2014} could be especially relevant because the secretions would be transported by the fluid flow.

In conclusion, our results  suggest that microbial range expansions on the surface of a highly viscous fluid provide a versatile laboratory system to explore the interplay between advection and spatial population genetics.

\begin{acknowledgments}
We would like to thank all members of A.W. Murray's group for their indispensable and generous help throughout this project.
Joanna Aizenberg's lab kindly allowing us to use their Kr{\"u}ss tensiometer to measure the surface tension of our fluid; we would especially like to thank Daniel Daniel and Michael Kreder for their time and helpful input. We would also like to thank Jennifer Lewis's and Dave Weitz's labs for allowing us to use their rheometers, and would like to particularly thank Sean Wei and Liangliang Qu for helping us optimize our measurements. S.A. and B.T.W. would like to thank Andrea Giometto for interesting discusions and his useful suggestions. We would also like to acknowledge conversations with Michael P.  Brenner. B.T.W. would like to thank Maxim Lavrentovich and Steven Weinstein for their helpful comments and guidance.
Work by B.T.W. was supported by the Department of Energy Office of Science Graduate Fellowship Program (DOE SCGF), made possible in part by the American Recovery and Reinvestment Act of 2009, administered by ORISE-ORAU under contract number\@ DE-AC05-06OR23100, by the US Department of Energy (DOE) under Grant No. DE-FG02-87ER40328. The Harvard MRSEC (DMR-1608501) helped to fund our usage of the Anton Paar rheometer. B.T.W., S.A., and D.R.N. benefitted from the National Science Foundation through grants DMR-1608501 and via the Harvard Materials Science and Engineering Center through grant DMR-1435999. S.A. and B.T.W. also benefitted from the National Science Foundation through DMS 1406870.
\end{acknowledgments}

\appendix
\counterwithin{figure}{section}
\counterwithin{table}{section}
\counterwithin{equation}{section}
\numberwithin{equation}{section}

\section{MATERIALS AND METHODS \label{sec:materials_and_methods}}

\subsection{Liquid substrate preparation}

To produce our highly viscous medium, standard rich growth medium for yeast (YPD), consisting of 1\% BactoYeast extract, 2\% BactoPeptone, and 2\% anhydrous dextrose (glucose), was mixed in autoclaved water and filtered into a sterile glass bottle using a Zapcap (Maine Manufacturing item number 10443430) to remove contaminants. We then systematically increased the substrate viscosity by adding 2-Hydroxyethyl cellulose, an extremely long-chain polymer with a viscosity-averaged molecular weight of $1.3 \times 10^6$ (Sigma-Aldrich product number 434981), at concentrations ranging from 2.0\%  to 3.0\% w/v into 300 mL aliquots of the media, as shown in Table \ref{tab:viscosity_table}. We used a strong magnetic mixer (IKA RCT basic magnetic stirrer) to rapidly stir the media with a sterile magnetic bar until it became homogenously viscous over the course of three hours.
We found that the model of the magnetic mixer was important; the mixer needed to be able to deliver enough torque to the stirbar so that it would continue spinning as the media became very viscous. Furthermore, if we used too much media in the mixing flask (typically volumes greater than 300 mL), the polymer would not mix evenly. The final mixture was sterilized to avoid contaminants brought in from the polymer. Because the extreme viscosity of our fluid prevented it from being filtered, we sterilized it by microwaving it for three minutes (with a ``Panasonic model number NN-SN9735'' microwave). In contrast to microwaving, sterilization via autoclaving produced inconsistent viscosities between replicates.
We found that it was \textit{essential} to let the media cool to room temperature in the bottle before pouring it into Petri dishes; yeast colony morphologies were not reproducible  when inoculated onto substrates prepared with different heating protocols. As discussed in Appendix  \ref{rheo}, the fluid's viscosity dropped almost 20\% over the first 24 hours and then slowly decreased as a function of time. The  cells were consequently always inoculated 24 hours after pouring the media. Future work should investigate how to make the fluid viscosity more stable.

\subsection{Strains}

We used the prototrophic (capable of synthesizing all required amino acids) yeast strains yJHK041 and yJHK042  which were derived from the W303 background. The two strains were virtually identical and differed only by the expression of different fluorescent proteins under the control of an ACT1 promoter. yJHK041 expressed mCitrine and was colored red in our figures while yJHK042 expressed mCherry and was colored cyan for visual clarity. yJHK041 had the genotype \textit{can1-100   bud4   prACT1-ymCitrine-tADH1-His3MX6:prACT1-ACT1} and yJHK042 had the same genotype except with \textit{ymCitrine} replaced with the \textit{ymCherry}. The two strains had identical growth rates in liquid culture and expanded at the same rate when deposited separately on agar plates.

\subsection{Standard experimental setup}

\begin{figure}
\centering
\includegraphics[width=8.6cm]{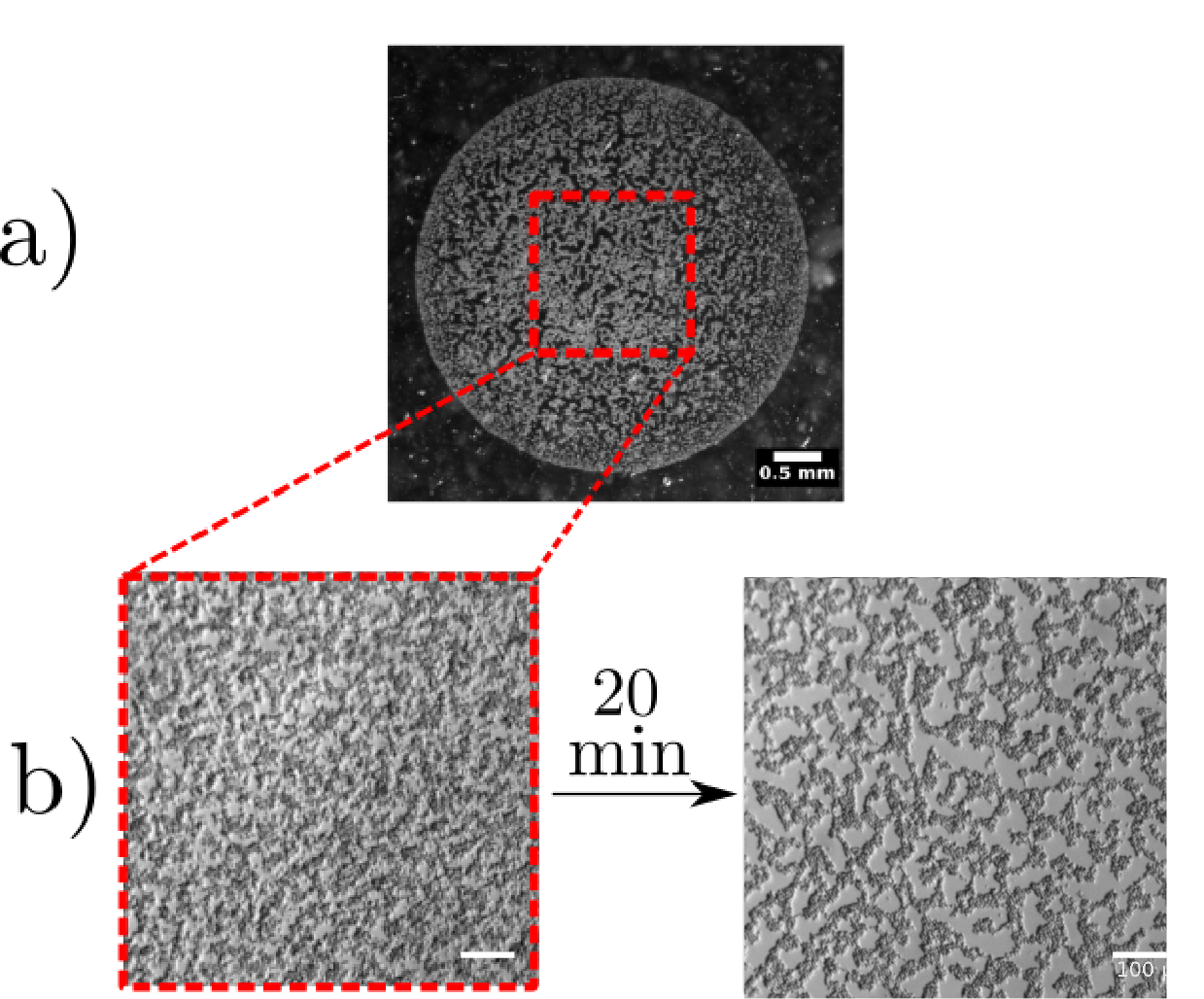}
\caption{
Upon deposition on the highly viscous substrate fluid, yeast cells first spread uniformly in the circular inoculant region usually called the ``homeland'' \cite{Hallatschek2007}, and then clump together via a coarsening process. a) Distribution of cells 20 minutes after inoculation on the viscous substrate, and b) zoom in immediately after inoculation (left) and after 20 minutes (right). The initially uniform distribution of yeast segregates into large clumps in a phase-separation process, suggestive of attractive interactions. The bottom scale bar corresponds to 100 $\mu$m.
\label{fig:yeast_spinodal_decomposition}}
\end{figure}

To prepare the saturated yeast cultures that we inoculated on our viscous media, we followed a similar procedure used for bacteria by Weinstein et al.\ \cite{Weinstein2017}. We took a single colony of yeast growing on an
agar plate and inoculated it in 10 mL of YPD media in a glass tube. The tube was then shaken overnight for roughly 16 hours at $30^\circ$ C as the yeast grew to saturation. The next morning, we used optical density measurements to place equal proportions of\@ yJHK041 and yJHK042 in an Eppendorf tube with a final volume of 1 mL.
After vortexing the Eppendorf tube, $2 \ \mathrm{\mu L}$ of saturated culture was taken from the tube and was inoculated on the surface of 40 mL\ of viscous fluid in a $94 \times 16$ mm Petri-dish (Greiner Bio-One item number 633181), leading to an average fluid height of $H = 7 \pm 0.2 \ \mathrm{mm}$. Throughout this paper, we used the same fluid height and volume unless specifically stated otherwise. Upon deposition, the cells immediately began to aggregate and formed clusters within 15 minutes as shown in Figure \ref{fig:yeast_spinodal_decomposition}. The plates were then wrapped with parafilm to inhibit drying and stored in a warm room held at $30^\circ$ C.

\subsection{Imaging \label{sec:imaging}}

The microbial colonies were imaged with an incubated Zeiss Lumar.V12 Stereoscope held at $30^\circ$C  with both fluorescent (eYFP and mCherry)\ and brightfield channels. In order to image large fields of view (i.e.\@ an entire Petri dish), we stitched many images together and blended their overlapping regions using Axiovision 4.8.2 software.
Our fluid was viscous enough that panning the microscope stage did not adversely shake the fluid and microbes.
Fluid flows were imaged by adding fluorescent green polyethylene microspheres between $10$ and $20 \ \mathrm{\mu m}$ in diameter (Cospheric item number ``UVPMS-BG-1.025 $10-20 \mathrm{um}$ - 0.1g'') before mixing the media with the polymer. We then imaged the position of the beads every 5 to 15 minutes, depending on the mean flow rate, with the eGFP channel. By varying the focal plane at which we observed the beads, we could follow the flow in a horizontal slice at the desired height, from the surface of the medium to the bottom of the Petri dish. The images were preprocessed and filtered before analyzing them with particle image velocimetry software (PIVlab for MATLAB), and the resulting velocity fields were post-processed using the Matlab tool PIVMat.

\subsection{Density measurements \label{sec:sophisticated_density_measurement}}

To test if yeast colonies depleted the density of the surrounding substrate as they metabolized, we compared the density of YPD media before and after the cells grew to saturation in it with an Anton Paar DMA 38 density meter.
To conduct this experiment, we placed a control test tube of YPD and another tube inoculated with our strains of yeast on a shaker overnight in a $30^\circ$C room; the yeast culture grew to saturation. The next day, we centrifuged both tubes, depositing the yeast on the bottom of the second tube, and measured the supernatant density of each.  We repeated this experiment three times and found that the average density of our control tube was  $\rho_{\text{YPD}} = 1.0167\pm 0.0003 \ \mathrm{g/mL} $ and that the density of the supernatant where the yeast had grown was $\rho_{\text{saturated}} = 1.0077 \pm 0.0003 \ \mathrm{g/mL}$, leading to a change in density of $\Delta \rho = -0.0090 \pm 0.0005  \ \mathrm{g/mL}$ where the $\pm$ corresponds to range of densities
that we measured.

\section{LIQUID SUBSTRATE RHEOLOGY \label{rheo}}

\begin{figure}
\centering
\includegraphics[width=8.6cm]{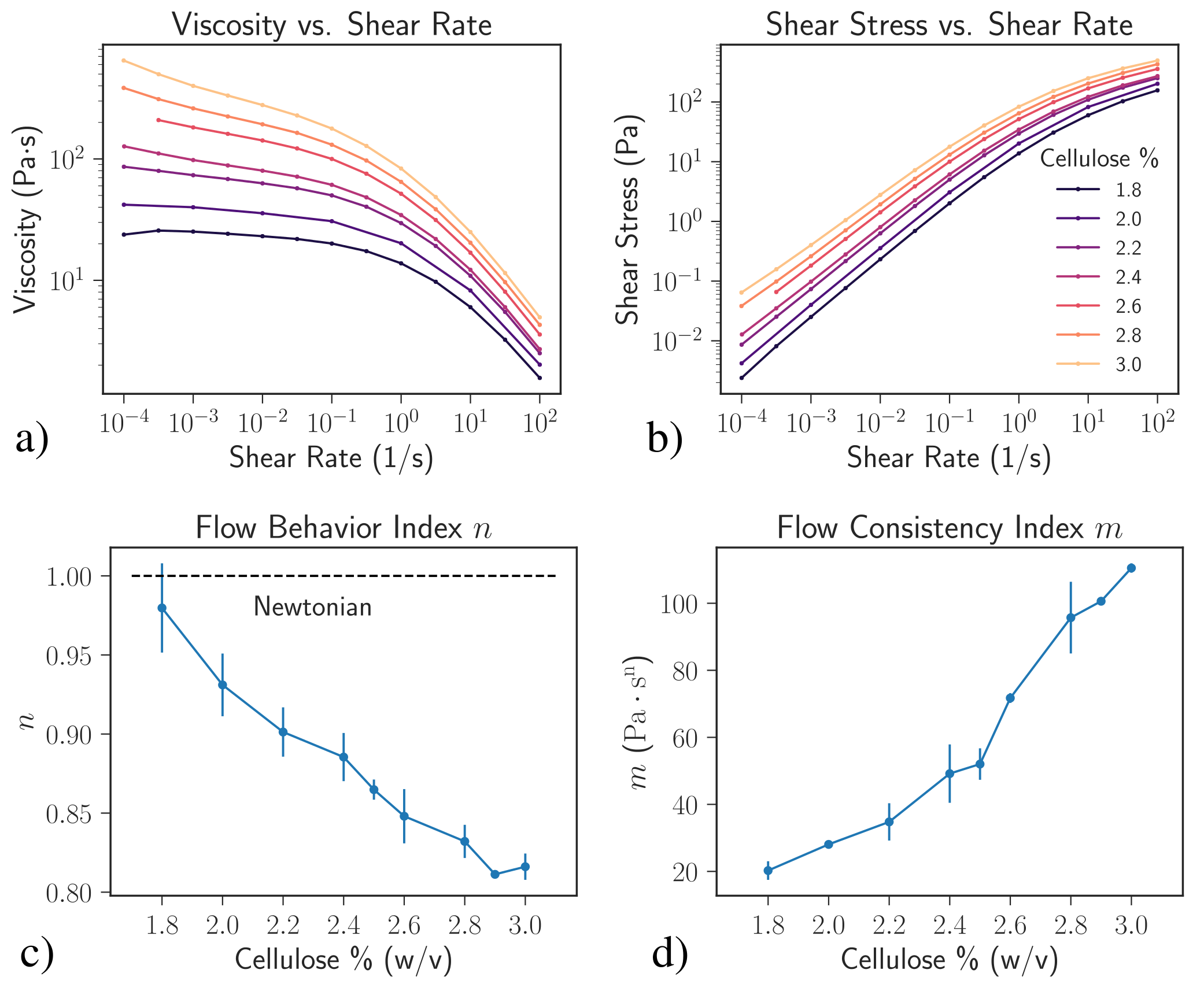}
\caption{(a) Shear viscosity and (b) corresponding shear stress for different polymer concentrations measured via steady-state flow tests. The
fluid was weakly shear thinning for $\dot{\gamma} \lesssim \ 10^{-1} \ \mathrm{s^{-1}}$
and reached a Newtonian plateau at less than or equal to 2\% polymer. For
other concentrations, a power law of $\tau=m \dot{\gamma}^n$ described the shear stress for shear rates of $\dot{\gamma} \lesssim \ 10^{-1} \ \mathrm{s^{-1}}$.  (c) and (d)\ plot $m$ and $n$ as a function of polymer concentration; the
fluid became more non-newtonian (shear thinning) as more polymer was added. \label{fig:rheology_varying_natrosal}
\label{fig:m_and_n_vs_cellulose}
}
\end{figure}

The substrate rheology was characterized with an Anton-Paar MCR 501 rheometer in a $50 \ \mathrm{mm}$ disk geometry with a $1 \ \mathrm{mm}$ gap. Fig. \ref{fig:rheology_varying_natrosal}a) displays steady-state flow tests for various polymer concentrations realized with logarithmic sweeps of the shear rate ranging from $10^2$ to $10^{-4}$ 1/s. Each point was averaged for several minute at $30^\circ  \mathrm{C}$ (the yeast incubation temperature), and the measurements were performed a day after the viscous media was microwaved, corresponding to the time that strains were inoculated on it.

Our viscous substrate exhibited a clear shear-thinning behavior, i.e.\ the viscosity decreased with increasing shear rate larger than $\dot{\gamma} \gtrsim 10^{-1}  \; \mathrm{s^{-1}}$ but presented a plateau for smaller shear rates. At cellulose concentrations higher than 2\%, the viscosity continued to decrease with shear rate for $\dot{\gamma} \lesssim 10^{-1} \ \mathrm{1/s}$ and we found an approximate power law relation between the shear stress $\tau$ and shear rate $\dot\gamma$; Figure \ref{fig:rheology_varying_natrosal}b) shows a fit to $\tau=m \dot{\gamma}^n$, in accord with the ``Power-Law''  model of Ostwald and de Waele \cite{bird_polymeric_liquids,Ostwald1925,de1923viscometry}, where the amplitude $m$ is the flow consistency index and the exponent $n$ corresponds to the flow behavior index.  The effective ``Newtonian'' viscosity of our fluid can then be expressed as $\eta\left(\dot\gamma \right)=m\dot{\gamma}^{n-1}$ \cite{bird_polymeric_liquids}, where $n=1$ describes Newtonian fluids and $n < 1$ indicates shear thinning behavior. We determined $m$ and $n$ as a function of polymer concentration by fitting the power law behavior at shear rates lower than $10^{-1} \ \mathrm{1/s}$ as shown in Figure \ref{fig:m_and_n_vs_cellulose}. Our liquid substrate exhibits increasing shear-thinning behavior (decreasing $n$) with larger polymer concentration; we found $n = 0.93 \pm 0.05$ at 2\% polymer and $n = 0.82 \pm 0.05$ at 3\%, suggesting a small, but measurable departure from Newtonian behavior across all polymer concentrations in this regime.

The typical shear rate in our experiments was on the order of $10^{-6} \leq \dot{\gamma} \leq 10^{-5} \; \mathrm{1/s}$, estimated from the measured surface flow velocity generated by the yeast colonies, $1 \leq u \leq 20$ mm/day, and with $\dot{\gamma} = u/H$ for a fluid with a typical height $H \approx 7$ mm. For simplicity, in this paper we described our substrate as a Newtonian fluid and determined the viscosity from its value at a shear rate $\dot \gamma = 10^{-4} \ \mathrm{1/s}$ (the lowest shear rate at which the rheometer give reproducible results); the corresponding values as we varied the polymer concentration are shown in Table \ref{tab:viscosity_table}. The media rheology was monitored over one week and presented a slow decrease in viscosity as a function of time after being microwaved (less than 10\% per day), and was neglected within the 3-5 day time scale of our experiments. Although we did not investigate closely, the viscosity of yeast complete synthetic media (CSM) appeared to be more stable as a function of time; future work should investigate this phenomenon.

\section{OTHER COLONY CONFIGURATIONS \label{sec:colony_orientation}}

\begin{figure}
\centering
\includegraphics[width=.48\textwidth]{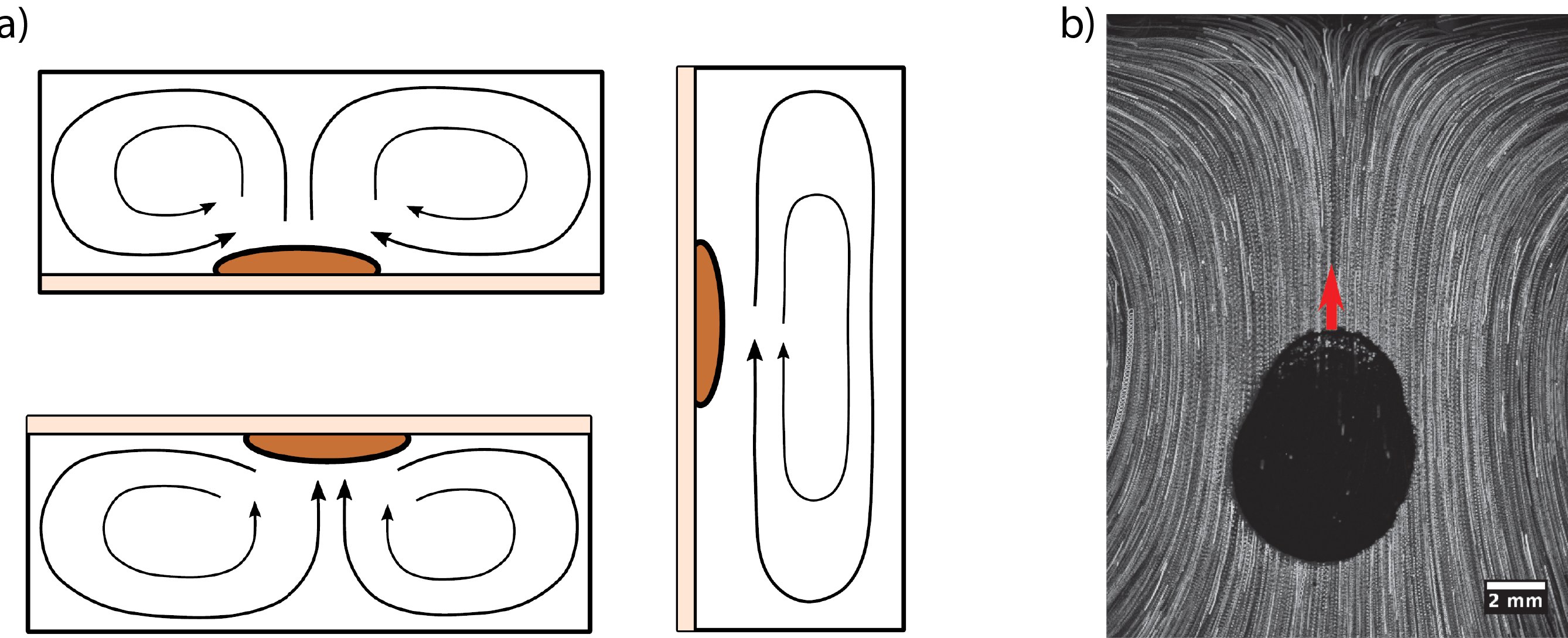}
\caption{\protect Fluid flow in sealed chambers with various yeast colonies' position relative to gravity. a) Schematic of a colony growing on a thin layer of agar on the bottom, top or side of a sealed container filled with our viscous media. Fluid flow near the colony, regardless of its position, is always generated opposing the direction of gravity. The fluid circulation was consistent in all cases with the vorticity direction predicted by baroclinic instability embodied in Eq. \eqref{eq:vorticity_equation}. b) same figure as in Fig. \ref{scheme}b) in the main text. Fluid flow streamlines corresponding to the yeast colony attached to the side. The streamlines were obtained by taking a maximum intensity projection of green fluorescent bead motion over several hours. The scale bar is 2 mm, and the red arrow indicates the direction of the flow for streamlines near the top of the colony. }
\label{fig:buoyant_flow_directions_and_plume}
\end{figure}

We explored other experimental geometries, summarized in Fig. \ref{fig:buoyant_flow_directions_and_plume}, to determine if other mechanisms ( e.g. Marangoni flows \cite{Scriven1960}) might account for the flow generated in the substrate fluids by the colony. However, experiments conducted where we anchored yeast colonies on a thin layer of agar to the top, bottom, and sides of sealed chambers filled with our viscous media Fig. \ref{fig:buoyant_flow_directions_and_plume}b) cast doubt on the Marangoni flow hypothesis. We found that colonies created fluid flows in the surrounding media similar in magnitude to experiments when the air-liquid interface was present regardless of their position in the chamber (even when placed at the top of the chamber), and also found that the induced fluid flows always opposed the direction of gravity. Although these experiments did not rule out the possibility that surface tension gradients drove flow when the free interface was present, they do suggest that metabolically-induced buoyant forces opposing the direction of gravity could be completely responsible for our observed flow. 

Buoyant flows result from differences in density in the presence of a gravitational field \cite{Turner1973} and, in our experiments, could originate from gradients in fluid temperature and solute concentration.
One possibility is that environmental temperature gradients (i.e.\ in the chamber where the yeast were imaged) drove fluid flows. As mentioned earlier, the very high viscosity of our liquid media substrates coupled with estimates of critical Rayleigh numbers strongly suggest that stray thermal gradients would be insufficient to produce convection in our experiments \cite{Benoit2008}. In fact, plates filled with viscous media, monitored over 24 hours, showed no evidence of a flow in the absence of yeast cells. The yeast colonies themselves must have induced buoyant flows by generating local gradients in the surrounding fluid's temperature or solute concentration. Similar to the work of Benoit et al.\ \cite{Benoit2008}, temperature gradients can be ruled out because heat diffusivity $D_\text{heat}$ is much larger than the molecular diffusivity $D_\text{glucose}$ of glucose in water, minimizing resulting density gradients caused by thermal gradients. This is can be estimated via the Lewis number of our media: $L = D_{\text{heat}} / D_{\text{glucose}} ~ 300$ indicative of an isothermal fluid. In addition, the coefficient of thermal expansion is much smaller than the coefficient of solute expansion; large temperature differences (several degrees Celsius) would be required to create the same density difference from a small change in solute concentration \cite{Benoit2008}. Estimates of the yeast cell metabolic heat production seem insufficient to produce the requisite thermal gradient. For instance, comparing the density change induced only by the cells glucose uptake $\Delta \rho_G$, with the density decrease due to the fluid thermal expansion caused by the heat produced during yeast glucose fermentation $\Delta \rho_T$, gives an estimate of $\Delta \rho_G / \Delta \rho_T \approx 1000$. This ratio suggests that the substrate density change is largely due to the glucose uptake rather than the metabolic heat produced by fermentation.

\section{CALIBRATING SIMULATION TO EXPERIMENTS \label{sec:calibrating_simulation}}

Table \ref{tab:model_parameters} in the main text shows the values used to fit our model to
experiment (i.e.\@  Figure \ref{simu_flow}), and the remainder of this appendix discusses how we obtained these values. 

\subsection{Viscous media density: $\rho_0$}

As discussed in section \ref{sec:sophisticated_density_measurement}, we found that the density of YPD media \textit{without} adding the cellulose polymer was $\rho_{\text{YPD}} = 1.0167\pm 0.0003 \ \mathrm{g/mL}$.
Mixing hydroxyethyl cellulose with water within the range of the concentration we used in our experiments, i. e. between 2\% and 3\%, did not significantly affect the solution density \cite{boutelier2016}. Additional density measurements of the polymer solutions when mixed with YPD solutions \cite{[{See also }]weinsteinThesis_2018} also didn't show a significant change in density of the substrate within experimental error.

Yeast colonies deplete the density of the surrounding media in order to create more biomass. In the model used by Benoit et al.\@ \cite{Benoit2008}), cells can absorb molecules with a variety of sizes and with correspondingly different concentration fields and diffusion constants. Here, the change in density we observed in overnight culture $\Delta \rho = -0.009 \pm 0.0004 \ \mathrm{g/mL}$ was consistent with approximatelly all of the glucose (originally 2\%) in the media being depleted within a factor of two \cite{crc}. For simplicity, we consequently used a single concentration field $c$ to model the diffusion and absorption of glucose only.

\subsection{Solute Expansion Coefficient: $\beta$} 

The solute expansion coefficient $\beta$ only enters in our dimensionless simulations via the combination $\beta c_1$ in the Rayleigh number,
\begin{equation*}
\mathrm{Ra}=\frac{h^3 \beta c_1 g}{D \nu},
\end{equation*}
where $c_1$ is the initial concentration of solute in the system. In our experiments, $c_1$ is also the \textit{maximum} concentration, since uptake of nutrients and excretion of less dense waste products leads to a net depletion of the effective concentration field. Hence, to estimate $\beta c_1$, we simply note that the density change when all solute is depleted is $\Delta \rho = -0.0090 \pm 0.0005 \ \mathrm{g/mL}$ from our experiments measuring the density of yeast overnight culture (Section \ref{sec:sophisticated_density_measurement}). Because the density of our media is $\rho = \rho_0 (1 + \beta c)$, and after a day of growth in well mixed culture, the glucose is completely depleted as the yeast can no longer reproduce, we estimate $\Delta \rho \approx - \rho_0 \beta c_1$, implying that $\beta c_1 = - \frac{\Delta \rho}{\rho_0}$.
After including appropriate sources of error, we thus find $\beta c_1=0.009\pm 0.001$. 
  
\subsection{Diffusion Constant: $D$}

\begin{figure}
\centering
\includegraphics[width=.4\textwidth]{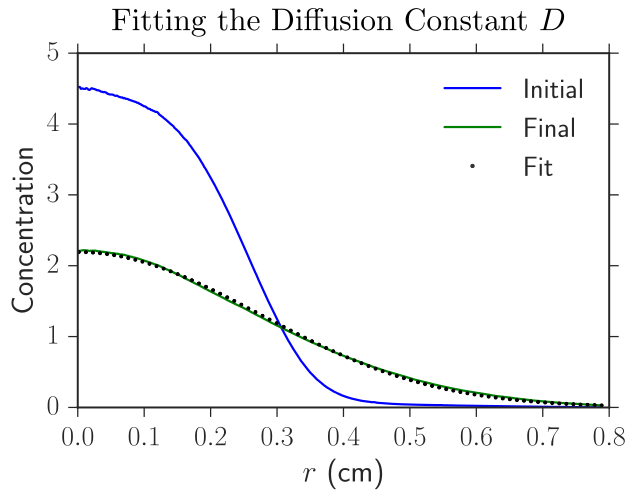}
\caption{
Fit to the fluorescein diffusion constant $D$ in our viscous fluid. The blue line corresponds to the measured initial radial profile, and the green line is measured the final profile.  Black dotted line represents the profile predicted by equation \eqref{eq:diffusion_constant_convolution} with the best fit value of $D$}
\label{fig:diffusion_constant_fit}
\end{figure}   

The glucose concentration field in our liquid substrate is difficult to track. In order to estimate the diffusion constant in our medium, we instead tracked the diffusion of fluorescein molecules as a proxy for glucose in our substrate over the course of several days (see \cite{[{See also }]weinsteinThesis_2018} for additional details). A circular droplet of approximately $7 \ \mathrm{mm}$ in diameter was deposited on the surface of a thin, $2.5 \ \mathrm{mm}$ thick layer of our viscous media. We used the Zeiss Lumar Stereoscope to confirm that the concentration of fluorescein was proportional to its fluorescent intensity at a fixed exposure time by creating a dilution series. We then imaged the droplet and extracted its radially symmetric concentration profile and repeated the process several hours later. The fluid was held at the same temperature as our colony expansion experiments ($30^\circ \mathrm{C}$). The radial density profile of a diffusing concentration $c(\vec{r},t)$ can be related to its original profile $c_{t_0} = c(r,t=0)$ via an integral representation that depends on the diffusion constant $D$ \cite{Debnath1995}:
\begin{equation}
\begin{aligned}
c(r,t) = \frac{1}{2 D t} \int_0^\infty ds \; s \; c_{t_0} I_0\left(\frac{rs}{2 D t}\right) e^{ - \frac{\left(r^2 + s^2\right)}{4D t}}
\label{eq:diffusion_constant_convolution}
\end{aligned}
\end{equation} 
where $I_0$ is the modified Bessel function of the first kind and we take the limit of the plate radius to infinity for simplicity. We ignored diffusion in the third dimension (towards the bottom of the plate) as the fluid layer was small relative to the droplet diameter. We fit the fluorescein diffusion constant $D$ in equation \eqref{eq:diffusion_constant_convolution} by inserting our experimentally measured initial concentration field at $c(r,t=0)$, numerically evaluating the integral, and comparing the predicted concentration field at later times to our experimental measurements. We adjusted the value of $D$ using least-squares to find the best-fit to our experimental measurement.

Figure \ref{fig:diffusion_constant_fit} displays the original radial profile of the fluorescein, the final profile, and the predicted fit from equation \eqref{eq:diffusion_constant_convolution} with the best value of $D$. We repeated this experiment three times on media with 2.0\%  and 3.0\% polymer concentration and found identical diffusion constants within experimental error, $D=2.4\pm 0.3 \; 10^{-6} \ \mathrm{cm^2/s}$. These results are consistent with the assumption that fluorescein diffusion is dominated by motion through the gaps between the long chains of hydroxyethyl cellulose polymer. Noting that the diffusion constants of fluorescein and dextrose (glucose) are similar in water at $25^\circ \mathrm{C}$: $D_\text{fluorescein}=4.25 \pm 0.01 \; 10^{-6} \ \mathrm{cm^2/s}$ and $D_\text{dextrose}=5.7 \; 10^{-6} \ \mathrm{cm^2/s}$ \cite{Culbertson2002} , for simplicity, we assumed that the nutrient diffusion constant in the substrate is similar to $D=2.4\pm 0.3 \; 10^{-6} \ \mathrm{cm^2/s}$ in our simulations of the substrate fluid.

\subsection{Mass flux rate into the yeast colony in rich nutrient conditions: $a c_1$ \label{sec:mass_flux_rate}}

\begin{figure}
\centering
\includegraphics[width=.4\textwidth]{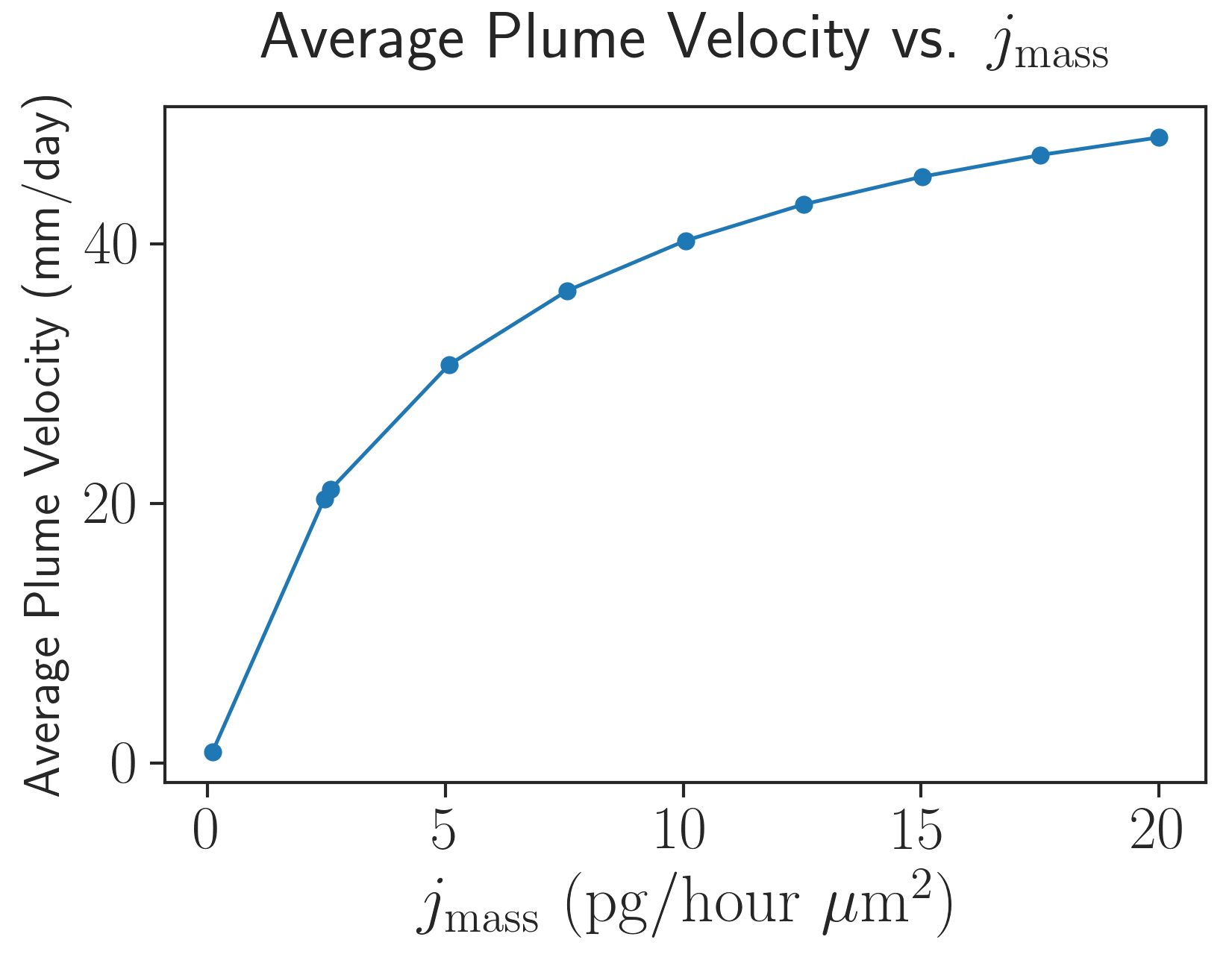}
\caption{
Simulated average flow velocity as a function of mass flux rate $j_\text{colony} = a c_1$ into a submerged yeast colony in rich nutrient conditions. The velocity field is determined above the center of the colony, in the rising plume of fluid from the bottom to the top of the domain.
}
\label{fig:fitting_jmass}
\end{figure}   

We fit $a c_1$, the mass flux rate into the yeast colony in rich nutrient conditions, by calibrating our simulation to experiments in a situation which negated the effect of surface tension: A yeast colony anchored on a thin agar sheet on the bottom of a \textit{sealed} petri dish filled with our viscous nutrient-containing fluid at $\eta = 54 \pm 8$ Pa$\cdot$s (see Figure \ref{fig:buoyant_flow_directions_and_plume}a); upper left). Under these conditions, the simulated yeast colony nutrient uptake created a buoyant plume in the direction opposing gravity, and the fluid flow reached a maximum, stable magnitude after about a day of growth. Note from the lower left side of Figure \ref{fig:buoyant_flow_directions_and_plume}a) that the induced flow in this case \textit{opposes} the outward growth-induced expansion velocity of the colony. We adjusted the product $a c_1$ until the simulated average flow velocity in the plume above the colony, as shown in Figure \ref{fig:fitting_jmass}, matched the average experimental velocity of tracer beads moving in the rising fluid from the bottom to the top of the container above the colony. The best match for $v_\text{experimental}=30 \pm 10 \ \mathrm{mm/day}$ resulted in a value of $a c_1 = 5 \pm 2 \ \mathrm{pg/(\mu m^2 \ hour)}$, where the $\pm$ is the standard deviation. 
 
We now argue that this mass flux rate is consistent with a simple, order of magnitude estimate and also show that it predicts a nutrient screening length inside yeast colonies in agreement with earlier investigations \cite{Lavrentovich2013}.

\paragraph{Order of magnitude estimate for $a c_1$}
A single yeast cell consumes about $N \sim 10^{12}$ glucose molecules per cell division when fermenting at high glucose concentrations \cite{,DeRisi1997}, and glucose has a molar mass of $M=180.156 \ \mathrm{g/mol}$. Yeast divide roughly every $\tau_g\approx90 \ \mathrm{minutes}$ in rich media, have a radius of approximately $r_\text{yeast}\approx 2.5 \ \mathrm{\mu m}$, and are approximately spherical when not actively dividing; they consequently have an area of $A_\text{yeast}=4\pi r_\text{yeast}^2$. Therefore, the glucose mass flux into a spherical yeast cell must be on the order of
\begin{equation} 
j_\text{cell} =a_\text{yeast} c\sim \frac{MN}{A_\text{yeast} \tau_g }\sim
 2.5 \ \frac{\text{pg}}{\mu\text{m}^2 \ \text{hour}}.
\end{equation}
In rich nutrient conditions, we assume that the concentration field is at its maximum value of  $c=c_1$ just outside the yeast cell walls, implying that $j_\text{cell} = a_\text{yeast} c_1$. Our order of magnitude estimate of $j_\text{cell}$ allows us then to estimate that $a_\text{yeast} c_1 \sim 2.5 \ \text{pg}/(\mu\text{m}^2 \ \text{hour})$, which is in the same order of magnitude as $a c_1$, the nutrient flux into the colony.

\paragraph{Consistency with nutrient screening length inside a yeast colony}

In the main text, we used our measured value of $a c_1$, the mass flux into the colony, to calculate the nutrient screening length in the fluid $\ell = (\rho_0 \beta D)/a = 5 \pm 2 \ \mathrm{mm}$. It is also possible to
use the value of $a c_1$ to estimate the nutrient screening length \textit{inside} the yeast colony given in \cite{Lavrentovich2013}:
\begin{equation}
\zeta = \sqrt{\frac{D \rho_\text{solute}}{\dot{\rho}}}
\end{equation}
where $\rho_\text{solute}=\rho_0 \beta c_1$ is the characteristic density of solute and $\dot{\rho}$ is the rate at which the solute is depleted. With the volume of a yeast cell $V_\text{yeast}=(4/3)\pi r_\text{yeast}^3$
and the packing fraction of spherical cells in a colony $\mathcal{N} \sim 0.5$, the value of $\dot{\rho}$ can then be estimated as:
\begin{equation}
\dot{\rho} \sim \mathcal{N} \frac{ j_{\mathrm{col} } A_\text{yeast}} {V_\text{yeast}}=  \frac{3 \mathcal{N} a c_1}{r_\text{yeast}}
\end{equation}
implying that the nutrient screaning length inside the colony is
\begin{equation}
\zeta = \sqrt{\frac{D \rho_0 \beta r_\text{yeast}}{3\mathcal{N} a} } \sim 90 \ \mathrm{\mu m},
\end{equation}
in approximate agreement with the work of Lavrentovich et al.\@ \cite{Lavrentovich2013}.

\section{NONDIMENSIONALIZING THE SET OF EQUATIONS \label{sec:nondimensionalizing_flow}}

As discussed in Sec. \ref{sec:model_of_dilation} of the main text, after coupling the Navier-Stokes equations with the diffusing solute field, applying the Boussinesq approximation as the local density variations are small in our experiments ($\delta \rho / \rho_0 \ll 1$), and including the flux boundary condition below the yeast colony, we find:
\begin{align}
\frac{\partial c}{\partial t} + \mathbf{u}\cdot \boldsymbol{\nabla} c &= D \nabla^2 c
\label{eq:diffusing_solute_with_velocity_appendix}
\\
\frac{\partial \mathbf{u}}{\partial t} + \mathbf{u}\cdot \boldsymbol{\nabla} \mathbf{u} &= - \frac{\boldsymbol{\nabla}
p}{\rho_0} + \nu \nabla^2 \mathbf{u} + \beta c\mathbf{g}
\label{eq:navier_stokes_boussinesq_appendix}
\\
\boldsymbol{\nabla} \cdot \mathbf{u} &= 0
\label{eq:incompressibility_appendix}
\\
\left(\boldsymbol{\nabla} c \cdot \mathbf{\hat{n}}\right)\bigg|_\text{colony} &= 
\left( \frac{c}{\ell} \right) \bigg|_\text{colony},
\label{eq:c_boundary_condition_near_yeast_appendix}
\end{align}
where $c$ is the nutrient concentration field, $\mathbf{u}$ the fluid velocity, $D$ the nutrient diffusion contant in the substrate, $\nu$ the fluid's kinematic viscosity, $\rho_0$ the substrate density without nutrient, $p$ the fluid's pressure, $\mathbf{g}$ the downward acceleration due to gravity, $\beta$ the solute expansion coefficient, $\mathbf{\hat{n}}$ the normal unit vector to the interface, and $\ell = \rho_0 \beta D / a$ the characteristic nutrient depletion length in the substrate fluid.

To better understand the dynamics of our model, we non-dimensionalize equations \eqref{eq:diffusing_solute_with_velocity_appendix}-\eqref{eq:c_boundary_condition_near_yeast_appendix} by choosing a characteristic length scale $L=H$, the height of the fluid in the Petri dish, a time scale $T=H^2/D$, (the time it takes solute to diffuse from the bottom to the top of the fluid in the petri dish), and the initial, \textit{maximum} glucose concentration $c_1$ (the initial concentration has the maximum value before the yeast cells deplete nutrients). The non-dimensionalized equations become:
\begin{align}
\frac{\partial \tilde c}{\partial \tilde t} + \mathbf{\tilde u} \cdot \boldsymbol{\nabla} \tilde c &= \nabla^2 \tilde c
\label{eq:diffusing_solute_dim}
\\
\frac{1}{\mathrm{Sc}}\left[\frac{\partial \mathbf{\tilde u}}{\partial \tilde t} + \mathbf{\tilde u} \cdot \boldsymbol{\nabla} \mathbf{\tilde u}\right] &
= - \boldsymbol{\nabla}\tilde p + \nabla^2 \mathbf{\tilde u} + \mathrm{Ra} \ \tilde c \mathbf{\hat{g}}
\label{eq:navier_stokes_boussinesq_dim}
\\
\boldsymbol{\nabla} \cdot \mathbf{\tilde u} &= 0
\label{eq:incompressibility_dim}
\end{align}
where the dimensionless concentration field is given by $\tilde c = c / c_1$, the dimensionless velocity is $\tilde u = u/(L/T)=u/(D/H)$ and the dimensionless pressure is $\tilde p =p / (D\rho_0\nu /H^2)$. 
The non-dimensional Navier-Stokes equation reveals two key dimensionless
parameters: the Schmidt number, $\mathrm{Sc} =  \nu/D$,
the ratio of the momentum diffusion to solute diffusion, and the Rayleigh number, $\mathrm{Ra}=(H^3 \beta c_1 g)/(D \nu)$ which quantifies the strength of the dimensionless buoyant force \cite{Guyon}. Non-dimensionalizing the flux boundary condition for the concentration field at the yeast colony's border reveals a final key parameter; the boundary condition becomes
\begin{equation}
\left(\boldsymbol{\nabla} \tilde c \cdot \mathbf{\hat{n}} \right)\big|_\text{colony} = \left( G \tilde c \right)\big|_\text{colony}
\label{eq:dimensionless_G_bc}
\end{equation}
where the ``mass flux number,'' $G = (H a)/(\rho_0 \beta D) \equiv H/\ell$
is the dimensionless ratio of the fluid height $H$ to the nutrient depletion length in the fluid $\ell$.  

The interplay between the Rayleigh, Schmidt, and mass flux numbers in our simulated geometry control the dynamics of our model. However, the large Schmidt number $\mathrm{Sc}=\nu/D \sim 10^8-10^9$ (using
the parameter values in Table \ref{tab:model_parameters}) allows us to set
the inertial terms in  equation \eqref{eq:navier_stokes_boussinesq_dim} to zero; this simplification corresponds to the Stokes regime.
Thus, we need only consider the interplay between the Rayleigh and mass flux numbers.
For the standard fluid height used in our experiments (40 mL of fluid in a standard $94$ mm diameter Petri dish, or $H = 7 \pm 0.2 \ \mathrm{mm}$), the Rayleigh number ranges
from $10^3$ to $10^4$ as we vary the fluid viscosity from $\eta=54\pm8$ Pa$\cdot$s to $\eta=600\pm90$ Pa$\cdot$s, and the mass flux number remains constant at $G \sim 4.4$. The yeast do not deplete nutrients quickly enough to allow us to set $c=0$ on the bottom of the colony, corresponding to the $G\rightarrow\infty$ limit. Both quantities consequently play a role in our experiments.

\section{SIMULATION METHODS\label{sec:simulation_methods}}

In this appendix, we discuss how we utilized OpenFOAM 5.0 \cite{openfoam} to simulate the buoyant fluid flow created by our yeast colonies and the early
stages of yeast colony growth.  Specifically, we discuss the particular programs that we created, how we prepared meshes and geometries for use in OpenFOAM, 
and how we analyzed and visualized simulation output. 

\subsection{\texttt{diffusionPressureFoam}}

The program \texttt{diffusionPressureFoam} (available on GitHub \cite{diffusionPressureFoam}) simulates a yeast colony that absorbs a diffusing concentration field and calculates the resulting hydrostatic pressure. We used \texttt{diffusionPressureFoam}
 to show how the baroclinic instability began before advection began to dominate, as seen in Fig. \ref{baro}. 

To create \texttt{diffusionPressureFoam}, we modified the standard solver packaged with OpenFOAM called \texttt{laplacianFoam}, which simulates a diffusing scalar field. At each timestep, we let the concentration field diffuse and possibly be absorbed
by the yeast. We utilized \texttt{swak4foam} \cite{swak4foam}, an extension of OpenFOAM, to impose the absorption boundary condition that
\begin{equation}
\left(\nabla c \cdot \mathrm{\hat{n}}\right)\big|_\text{colony} = \left( \frac{a c}{\rho_0 \beta D} \right) \bigg|_\text{colony}.
\end{equation}
The hydrostatic pressure inside a fluid is given by \cite{Guyon}
\begin{equation}
\boldsymbol{\nabla} p = \rho_0(1+\beta c) \mathbf{g}.
\label{eq:hydrostatic_pressure}
\end{equation} 
However, to calculate the hydrostatic pressure numerically, we took the divergence of equation \eqref{eq:hydrostatic_pressure} and solved
\begin{equation}
\nabla^2 p = \boldsymbol{\nabla} \cdot \left[\rho_0(1+\beta c) \mathbf{g}\right].
\end{equation}
At the free interface, we imposed the boundary condition that $p=p_\text{atmospheric}$
while on other walls, we imposed the condition (again using \texttt{swak4foam} \cite{swak4foam}) that
\begin{equation}
\left[\boldsymbol{\nabla} p \cdot \mathrm{\hat{n}} \right]\big|_\text{walls} 
= 
\left[\rho_0(1+\beta c) \mathbf{g} \cdot \mathrm{\hat{n}} \right] \big|_\text{walls}.
\end{equation}

We always assumed radial symmetry when simulating yeast colonies on the surface of a viscous nutrient-containing liquid or at the bottom of a sealed petri dish. To create our radially symmetric geometry, we used \texttt{gmsh} 3.0.5 \cite{Lee2011} to create a 2-dimensional structured mesh spanning the petri dish and then
extruded it to form a wedge with an angle of $2.5^\circ$ which is the appropriate
setup for radially symmetric simulations in OpenFOAM. We simulated our experiments
at a resolution such that 20 simulation cells spanned the yeast colony radius. We wrapped the \texttt{gmsh} geometry
creation script in python scripts that could automatically generate geometries,
change simulation parameters, and quickly analyze simulation output.

\begin{figure}
\centering
\includegraphics[width=8.6cm]{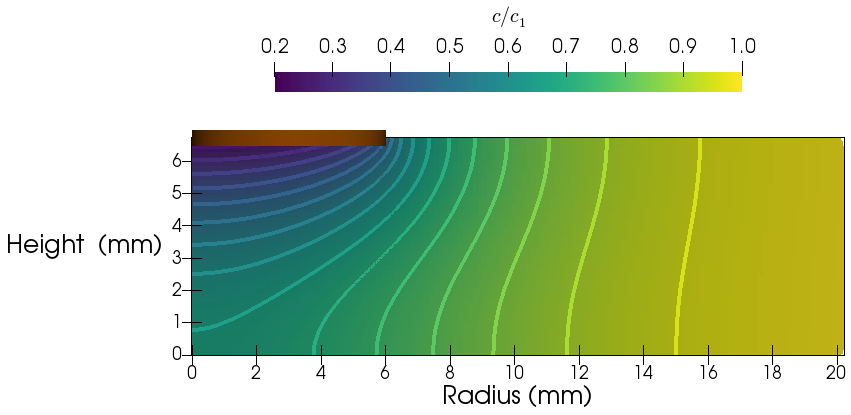}
\caption{Radially symmetric simulation of the concentration field $c/c_1$ below a yeast colony (the thick brown bar) on the surface of our viscous liquid after 48 hours
and at $\eta=400 \pm 50 \ \mathrm{Pa \cdot s}$. Equally spaced contours of constant concentration are shown.
\label{fig:simulation_cfield_and_ufield}}
\end{figure}

After running a simulation, we used Paraview \cite{Ayachit2015}, an open-source tool to visualize large geometrical datasets, to visualize the results and create figures such as the concentration field showed in Fig. \ref{fig:simulation_cfield_and_ufield}. To quickly analyze data from many simulations, we used automated Python scripts to extract relevant data such as the velocity on the fluid's surface and the total amount of solute present in a petri dish. To create the baroclinicity field $\frac{1}{\rho^2}\left(\boldsymbol{\nabla} \rho \times \boldsymbol{\nabla} p \right)$ seen in Fig. \ref{baro}, we utilized the \texttt{funkySetFields} utility, a part of \texttt{swak4foam} \cite{swak4foam}, which can algebraically manipulate the output of OpenFOAM simulations.

\subsection{\texttt{stokesBuoyantSoluteFoam}}

The program \texttt{stokesBuoyantSoluteFoam}  simulated
how yeast depleted the density of the surrounding fluid and calculated the resulting
fluid flow. We used this program to generate the quantitative agreement
between experiment and simulation in Figure \ref{simu_flow}. Specifically, \texttt{stokesBuoyantSoluteFoam} solves
the dimensionless equations \eqref{eq:diffusing_solute_dim}-\eqref{eq:incompressibility_dim}
and the dimensionless mass flux boundary condition below the yeast colony
in equation \eqref{eq:dimensionless_G_bc}. It assumes that the Schmidt number $S_c=\nu/D$ is infinite (as discussed above) and consequently solves
\begin{align}
\frac{\partial \tilde c}{\partial \tilde t} + \mathbf{\tilde u} \cdot \boldsymbol{\nabla} \tilde c &= \nabla^2 \tilde c
\\
0 &= -\boldsymbol{\nabla} \tilde p + \nabla^2 \mathbf{\tilde  u} + \mathrm{Ra} \ \tilde c \mathbf{\hat{g}}
\\
\boldsymbol{\nabla} \cdot \mathbf{\tilde u} &= 0
\\
\left(\boldsymbol{\nabla} \tilde c \cdot \mathbf{\hat{n}} \right)\big|_\text{colony} &= \left( G \tilde c \right)\big|_\text{colonys}
\label{eq:mat_meth_fluxBC}
\end{align}
at each timestep. 

We again utilized \texttt{swak4foam} \cite{swak4foam} to implement
the concentration boundary condition at the yeast colony boundary (equation \ref{eq:mat_meth_fluxBC}).   At
each timestep, the solute diffused and was absorbed by the yeast. After diffusing,
\texttt{stokesBuoyantSoluteFoam} calculated the steady-state velocity field using the same technique as \texttt{buoyantBoussinesqSimpleFoam} (the SIMPLE algorithm \cite{Ferziger2002}) which is packaged with OpenFOAM.
The velocity from the previous timestep was used as an initial guess for the velocity field in the next time step to improve its convergence speed. To avoid stability problems resulting from a high Courant number \cite{Ferziger2002}, we adaptively changed the timestep  to ensure that the maximum Courant number
in the simulation remained below 0.5  and also used the implicit Crank-Nicolson technique \cite{Ferziger2002} to evolve the concentration field. Geometry preparation and postprocessing for \texttt{stokesBuoyantSoluteFoam} was
the same as that for \texttt{diffusionPressureFoam}.

\subsection{\texttt{forcedThinFilmFoam}}

\begin{figure}
\includegraphics[width=.4\textwidth
]{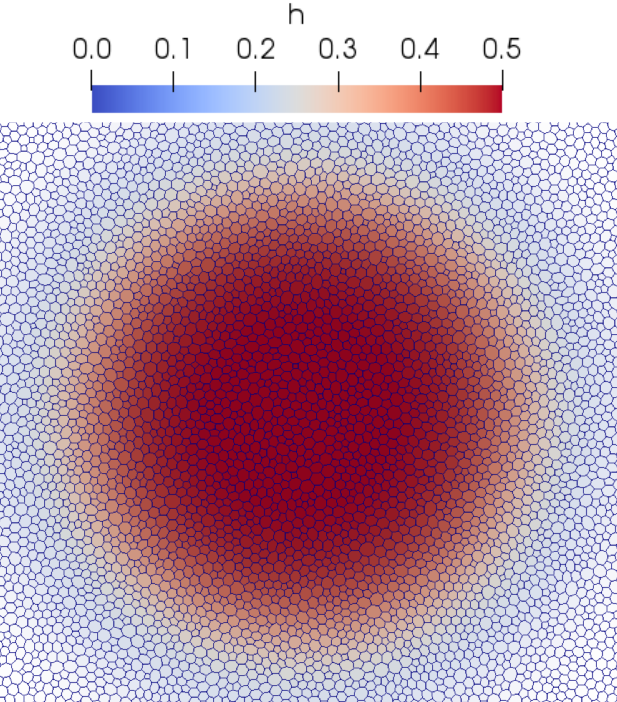}
\centering
\caption{
A polyhedral dual mesh (blue cells) and corresponding height field
$h$ used by the \texttt{forcedThinFilmFoam} program \cite{forcedThinFilmFoam}.
\label{fig:forcedThinFilmFoam_polyhedral_mesh}}
\end{figure}

\texttt{forcedThinFilmFoam} (available on GitHub \cite{forcedThinFilmFoam}) solves equation \eqref{model} in the main text, or
\begin{equation}
\begin{aligned}
\frac{\partial h(\mathbf{r},t)}{\partial t}     &+ \boldsymbol{\nabla} \cdot \left[ h(\mathbf{r},t) \mathbf{v}(\mathbf{r}) \right]\\
                                                                        &= D_h \nabla^2 h(\mathbf{r},t) + \mu h(\mathbf{r},t) \left[ 1 - \frac{h(\mathbf{r},t)}{h_0} \right].
\label{eq:model_appendix}
\end{aligned}
\end{equation}
and leads to the radial height profiles shown in Fig. \ref{h_simu}. Although we could simulate arbitrary velocity fields, we used
the radially symmetric field of $\mathbf{v}(\mathbf{r})=(1/2)\alpha r \ \hat{\mathbf{r}}$,
matching equation \eqref{eq:radial_velocity_and_alpha}.

We used \texttt{gmsh} 3.0.5 \cite{Lee2011} to create a two-dimensional mesh in a circular
domain mimicking a Petri dish as seen in Figure \ref{fig:forcedThinFilmFoam_polyhedral_mesh}. We found that the choice of mesh \textit{dramatically} impacted simulation performance; using a regular cartesian grid led to pronounced
lattice artifacts, likely because of the autocatatalytic growth term on
the right side of equation \eqref{eq:model_appendix}. We obtained the best
results when we converted a Delaunay  triangular mesh to its dual polyhedral mesh using OpenFOAM's \texttt{polyDualMesh} utility,
similar to other work simulating fluid flow in radial geometries \cite{Vita2012a}.

To allow for advection-dominated simulations, we used a flux-limiting Super bee scheme when calculating the divergence term, or $\boldsymbol{\nabla} \cdot \left[ h(\mathbf{r},t) \mathbf{v}(\mathbf{r}) \right]$. To prevent stability problems, we ensured that the maximum Courant number was less than 0.1 and  used the implicit Crank-Nicolson technique \cite{Ferziger2002} to evolve the height field. We again utilized Python scripts to analyze the data coupled with OpenFOAM's postprocessing \texttt{singleGraph} tool.

\section{Simulated Nutrient Absorption vs. Flow Rate \label{sec:nutrient_absorption_buoyant_flow}}

\begin{figure}
\includegraphics[width=8.6cm
]{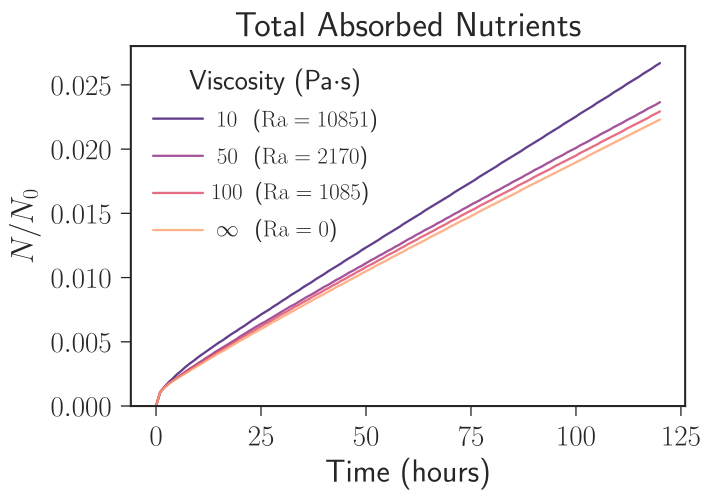}
\centering
\caption{
Total number of nutrient molecules absorbed by the yeast relative to the original number of nutrient molecules in the fluid $N/N_0$ as the fluid viscosity is varied. The height of the fluid is $H=7 \ \mathrm{mm}$ and the rest of the simulation parameters are set using the values in Table \ref{tab:model_parameters}. The Rayleigh number, $ \mathrm{Ra}=(H^3 \beta c_1 g)/(D \nu)$,  varies from $0$ to approximately $10^4$ due to the changing viscosity $\nu$, and the mass flux number, $G = (H a)/(\rho_0 \beta D) \equiv H/\ell$, is fixed at $G \approx 4.4$. Note that the stronger advection of the substrate fluid at lower viscosities leads to an enhenced uptake of nutrient molecules.
\label{fig:mass_absorption_vs_time}}
\end{figure}

To investigate if microbial colonies generating buoyant flows absorb more nutrients than those that do not, we simulated a yeast colony on the surface of our fluid (again with a fixed colony radius for simplicity) and varied the substrate viscosity, from $10$ Pa$\cdot$s to $100$ Pa$\cdot$s, allowing us to control the magnitude of the buoyant flow. We also simulated a substrate with infinite viscosity where no flow was allowed. We kept the rest of the simulation parameters fixed to the values in Table \ref{tab:model_parameters} with $H=7 \ \mathrm{mm}$ and recorded the nutrient uptake by the colony over time. The Rayleigh number, $\mathrm{Ra}=(H^3 \beta c_1 g)/(D \nu)$, of these simulations ranged between $0$ and $10^4$ and the mass flux number, $G = (H a)/(\rho_0 \beta D) \equiv H/\ell$, was fixed at $G \approx 4.4$. As shown in Figure \ref{fig:mass_absorption_vs_time}, the more vigorous flows associated with smaller substrate viscosities allowed yeast colonies to absorb nutrients more efficiently; the nutrient absorption rate at $10$ Pa$\cdot$s was about 1.5 times larger than at infinite viscosity.
It is possible that microbes growing on less viscous fluids could induce more intense flows, enhancing this effect even further.  It thus seems plausible that colonies generating stronger buoyant flows could indeed have a selective advantage.

\bibliography{library}
\addcontentsline{toc}{section}{Bibliography}

\end{document}